\documentclass[aps,twocolumn,nofootinbib,superscriptaddress,amsfonts,floatfix
]{revtex4-1} % for review and submission
\usepackage{graphicx,amsmath,amssymb,amstext}
\usepackage{amssymb,amsbsy,amsfonts,amsthm,color}

\usepackage{mathtools,nccmath}

\usepackage{epsfig}
\usepackage{graphicx}
\usepackage{subfigure}
\usepackage[dvipsnames]{xcolor}
\definecolor{linkcolor}{rgb}{0.0,0.3,0.5}
\usepackage[unicode, colorlinks=true, linkcolor=linkcolor, citecolor=linkcolor, 
filecolor=linkcolor,urlcolor=linkcolor, pdfusetitle]{hyperref}
\usepackage{cancel}
\usepackage{balance}
\usepackage[capitalise]{cleveref}
\usepackage{xspace}
\usepackage{enumerate}
\usepackage{ulem}
\usepackage{mdframed}
\usepackage{booktabs}
\usepackage{makecell}
\usepackage{enumitem,tabularx,multirow}
\usepackage{aas_macros}
\usepackage{colortbl}

\AtBeginDocument{%
  \heavyrulewidth=.08em
  \lightrulewidth=.05em
  \cmidrulewidth=.03em
  \belowrulesep=.65ex
  \belowbottomsep=0pt
  \aboverulesep=.4ex
  \abovetopsep=0pt
  \cmidrulesep=\doublerulesep
  \cmidrulekern=.5em
  \defaultaddspace=.5em
}

\graphicspath{{Figures/}}

\usepackage{color}

\newcommand{\Beq}{\begin{eqnarray}}
\newcommand{\Eeq}{\end{eqnarray}}

\newcommand{\nn}{\nonumber \\}

\def\lsim{\mathrel {\vcenter {\baselineskip 0pt \kern 0pt \hbox{$<$} \kern 0pt \hbox{$\sim$} }}}

\def\gsim{\mathrel {\vcenter {\baselineskip 0pt \kern 0pt \hbox{$>$} \kern 0pt \hbox{$\sim$} }}}

\newcommand{\RomanNumeralCaps}[1]

\definecolor{mypurple}{RGB}{143, 116, 210}

\def\-{\,-\,}
\def\={\,=\,}
\def\+{\,+\,}
\def\equi{\,\equiv\,}

\definecolor{burgundy}{rgb}{0.5, 0.0, 0.13}
\definecolor{coolblack}{rgb}{0.0, 0.18, 0.39}
\definecolor{darkblue}{rgb}{0.0, 0.0, 0.55}
\definecolor{darkgreen}{rgb}{0.0, 0.2, 0.13}

\newcommand{\jhu}{Department of Physics and Astronomy,
Johns Hopkins University, Baltimore, MD 21218, USA}

\begin{document}
\title{Cavity effect in the quasinormal mode spectrum of topological stars}

\author{Pierre Heidmann}
\email{pheidma1@jhu.edu}
\affiliation{\jhu}

\author{Nicholas Speeney}
\email{nspeene1@jhu.edu}
\affiliation{\jhu}

\author{Emanuele Berti}
\email{berti@jhu.edu}
\affiliation{\jhu}

\author{Ibrahima Bah}
\email{iboubah@jhu.edu}
\affiliation{\jhu}

\begin{abstract}
We study scalar perturbations of topological solitons, smooth horizonless solutions in five-dimensional Einstein-Maxwell theory that correspond to coherent states of gravity via the dynamics of extra compact dimensions. First, we compute scalar quasinormal modes for topological stars that have a single unstable photon sphere, and we show that the spectrum is very similar to that of a black hole with the same photon sphere. Next, we study topological stars that have both a stable inner photon sphere and an unstable one. The first few quasinormal modes are localized around the inner photon sphere. The spectrum also contains ``black-hole like modes'' localized at the unstable outer photon sphere. The frequencies of these modes are similar to those of a black hole, but their imaginary part is smaller due to a cavity effect associated with the inner photon sphere. The longer damping produced by this trapping effect may have implications for black hole spectroscopy. 

\end{abstract}
%\pacs{}
\maketitle

%%%%%%%%%%%%%%%%%%%%%%%%%%%%%%%%%%%%%%%%%%%%%%%%%%%%%%%%%%%%%%%%%%

\section{Introduction} \label{sect:intro2}

Direct measurements of gravitational waves have opened a new observational window for testing general relativity. They offer compelling prospects for probing the strong gravity environment near black holes, and the equally exciting possibility of observing exotic compact objects beyond general relativity~\cite{LISA:2022kgy}. These observational developments could allow for a deeper exploration of the quantum aspects of gravity.

In quantum gravity, black holes correspond to thermodynamic ensembles of quantum states. The general paradigm necessary to fully characterize such states is still lacking. However, a subset of them can be coherent enough to admit classical descriptions. Many examples of such states can be constructed from string theory and characterized in various theories of gravity (see Refs.~\cite{Lin:2004nb,*Kanitscheider:2006zf,*Giusto:2004ip,*Bena:2022ldq,*Bena:2022rna,*Warner:2019jll} for a non-exhaustive list). 

These coherent states lead to smooth horizonless geometries, generated by gravitational solitons in spacetime and supported by electromagnetic flux.  For many years, they could only be obtained from supersymmetric theories of gravity, and thereby beyond what could be considered astrophysically relevant (see Refs.~\cite{Bena:2020yii,Bena:2020see,*Bacchini:2021fig,*Bianchi:2020bxa,*Bah:2021jno,*Mayerson:2020tpn,*Bianchi:2020yzr,*Bianchi:2020des,*Bianchi:2021xpr,*Bianchi:2021mft,*Bianchi:2022qph,*Ganchev:2022vrv,Ikeda:2021uvc} for some analysis of their gravitational signatures).

Two of the authors provided new mechanisms for constructing smooth and horizonless solutions in generic, non-supersymmetric theories of gravity with compact extra dimensions.  These solutions are gravitational solitons induced by nontrivial topological microstructures in the internal space and supported by electromagnetic flux~\cite{Bah:2020ogh,Bah:2020pdz,*Bah:2021owp,*Bah:2021rki,*Bah:2022yji,*Bena:2022tro,*Bah:2023ows,Heidmann:2021cms,*Bah:2022pdn,*Heidmann:2022zyd}.  Moreover, the solutions can be embedded in string theory, and appropriately interpreted as coherent states of quantum gravity~\cite{Heidmann:2021cms,*Bah:2022pdn,*Heidmann:2022zyd}. They are referred to as \emph{topological solitons}, and can be seen as a new topological phase of matter that is inherently geometric.    

In Ref.~\cite{Heidmann:2022ehn}, some of the authors investigated the gravitational signatures of topological solitons by characterizing the general phenomenology of photon geodesics, including the properties of lensing around these objects. In this paper, we initiate a study of their quasinormal modes (QNMs). This will bring us one step closer to understanding their possible relevance in describing observational alternatives to black holes from quantum gravity.  
More specifically, we will study scalar QNMs of the simplest topological solitons: the {\it topological stars}, which correspond to spherically symmetric static spacetimes consisting of a single charged Kaluza-Klein bubble~\cite{Bah:2020ogh}.

The QNMs are observables encoded in gravitational-wave signals of mergers of compact objects. 
The signals themselves consist of three characteristic phases (inspiral, merger, and ringdown), and new physics can show up in any of them if there are deviations from general relativity.  The ringdown phase, in particular, is dominated by the QNMs, which are deformations of the geometry characterized by superpositions of exponentials with complex frequencies, i.e., damped oscillators.  They can be computed by studying linear perturbations of the final state.  In this paper, we will obtain the QNMs corresponding to scalar perturbations of topological stars, and compare them with the  QNM spectrum of black holes.  

Single-bubble topological stars are not particularly relevant for describing astrophysical states of gravity because they carry a magnetic charge, and they behave as ``space-time mirrors'' for photon scattering~\cite{Heidmann:2022ehn}. However, they have several desirable features that make them excellent prototypes for exploring phenomenological aspects of topological solitons in general.  

First, their linear perturbation equations are separable. This allows us to compute QNMs and their eigenfunctions using standard techniques.  As we will see below, these QNMs exhibit interesting phenomena that we expect to be present in generic topological solitons.  

Second, the solitons do not admit horizons, and yet they are compact enough that they can have one or two photon spheres, which can be stable or unstable.  This allows us to directly investigate the relation of QNMs with properties of these surfaces, and in particular, sharply determine which aspects of QNMs are fixed by horizons or by photon spheres.\footnote{This interesting point has been discussed in the context of exotic horizonless compact objects (see e.g.~\cite{Cardoso:2016rao,Cardoso:2016oxy,Cardoso:2019rvt}). The main conclusions are that (i) the postmerger ringdown waveform of exotic, horizonless ultracompact objects is initially identical to that of a black hole, and (ii) putative corrections at the horizon scale will appear as secondary pulses (``echoes'') after the main burst of radiation. However, in four spacetime dimensions, the presence of a stable photon sphere is generically associated with a nonlinear instability~\cite{Keir:2014oka} -- different from the ``usual'' superradiant ergoregion instability that affects rotating ultracompact objects without horizons~\cite{1978CMaPh..62..247F,1978RSPSA.364..211C} -- that can destabilize these objects on a dynamical timescale~\cite{Cardoso:2014sna,Cunha:2017qtt,Cunha:2022gde}. The instability occurs quite generically but it relies on a crucial assumption on topology -- i.e., that the spacetime is continuously deformable into Minkowski.}
We are also able to compare the properties of QNM spectra coming from systems with a varying number of photon spheres.  This is interesting since one can expect novel interference patterns 
related to the so-called gravitational-wave ``echoes.''

Static black holes have an unstable photon sphere, commonly referred to as the ``shadow,'' surrounding the event horizon. Their QNMs can be broadly classified into two categories: the \emph{slowly damped modes} and the \emph{highly damped modes}~\cite{Schutz:1985km,Iyer:1986np,Leaver:1985ax,*Leaver:1990zz}. Modes in the first category correspond to perturbations localized at the photon sphere and are governed by the instability properties of the corresponding orbits~\cite{Cardoso:2008bp}. The real part of their frequencies (which would correspond to normal frequencies in a self-adjoint problem) is proportional to the angular velocity of photons. The imaginary parts (associated with the damping time of the perturbations) are usually much smaller, and they are related to the instability timescale of the photon orbit, which is given by the Lyapunov exponent. In the eikonal (large angular momentum) limit, the QNM frequencies are given by
\begin{equation}
    \omega_N^\text{BH} \,\sim\, \Omega \, (\ell+\tfrac{1}{2}) + i \,\lambda (N+\tfrac{1}{2})\,,
    \label{eq:IntroEq}
\end{equation}
where $\Omega$ and $\lambda$ are the angular velocity and Lyapunov exponent at the photon sphere, while $\ell$ and $N$ are the spherical harmonic index and the so-called ``overtone number,'' respectively. For large $N$, the mode decay time scale can become comparable to their period or faster. Modes with $\text{Im}(\omega_N) \gtrsim \text{Re}(\omega_N)$ 
are highly damped, and they are not expected to contribute significantly to the signal. 

In this paper, we want to understand whether the broad features of the QNM spectrum are modified for topological solitons. Just like black holes, topological solitons are characterized by an outer unstable photon orbit~\cite{Heidmann:2022ehn}. However, their inner structure does not possess an event horizon, and therefore Eq.~\eqref{eq:IntroEq} may not apply as it stands. 

Topological stars are particularly interesting solutions to understand modifications of the QNM spectrum, because they can be of two types. The first type has a single photon sphere, which is unstable; the second type has {\em both} a stable and an unstable photon sphere. 

First, we show that topological stars with a single unstable photon sphere have the same structure of QNMs as in Eq.~\eqref{eq:IntroEq}, while having no horizon.  This confirms that the QNMs of black holes capture properties of the unstable photon sphere, rather than properties of the horizon (see e.g.~\cite{1972ApJ...172L..95G,Mashhoon:1985cya,Berti:2005eb,Berti:2009kk,Cardoso:2008bp,Cardoso:2016rao,Cardoso:2019rvt}).

Second, the topological stars with a pair of photon spheres have a much richer mode structure.  There are different time scales associated with the two shells.  Moreover, the structure of the QNMs can be qualitatively different due to the stability properties of the photon rings.  In particular, we noticed that the QNM spectra have two important properties:

\begin{itemize}
    \item[(1)] \underline{Echoes at late times.}
\end{itemize}
The first few fundamental modes are localized at the inner stable photon orbit, and their dynamics are governed by the scattering properties there. Their normal frequencies are small, 
and the presence of multiple time scales leads to phenomena similar to echoes at late times~\cite{Bena:2019azk}.
Moreover, their imaginary frequencies are exponentially suppressed as a function of $\ell$ so that their damping time is very large, which is a characteristic of waves trapped on a stable surface.

\begin{itemize}
    \item[(2)] \underline{Black hole modes at early times with a cavity effect.}
\end{itemize}
As we increase the overtone number, $N$, the modes start to get localized at the outer unstable photon orbit. Here, their features are determined by the instability properties of geodesics, and they are very similar to those of a black hole with the same photon sphere.  More precisely, they have the same normal frequencies as the black hole modes. However, their imaginary frequencies differ, and they have a longer damping time compared to the QNMs of black holes. This suggests that the short-term gravitational-wave signal from a topological soliton would have the same frequency as black holes, but with {\it a longer damping time}.  This phenomenon is due to a cavity effect, produced by a nontrivial interaction between stable and unstable orbits that does not exist in the presence of an event horizon.
\\

These two properties are expected to generically characterize any compact object with a stable photon sphere and an unstable photon sphere.  In particular, we expect them to be present in generic horizonless and smooth topological solitons.  Echoes induced by inner stable orbits have been largely studied in various horizonless ultracompact objects, including neutron stars~\cite{Cardoso:2014sna}, gravastars~\cite{Cardoso:2016oxy}, boson stars~\cite{Cardoso:2016oxy}, traversable wormholes~\cite{Cardoso:2016rao,Dimitrov:2020txx}, or supersymmetric topological solitons~\cite{Bena:2019azk,Bena:2020yii,Ikeda:2021uvc}. However, the presence of damping differences or cavity effects (to our knowledge) has not been pointed out before, and it could lead to novel, interesting phenomenological implications for future experiments.

Before proceeding, we provide a roadmap for the paper and a summary of our main results. In Secs.~\ref{sec:TSGen} and \ref{sec:ScalWave} we review the properties of topological stars and derive the equations governing scalar perturbations. In Sec.~\ref{sec:WKB} we develop generic Wentzel-Kramers-Brillouin (WKB) methods to compute the spectrum of slowly damped QNMs for smooth horizonless geometries. In Sec.~\ref{sec:TSQNM} we apply those methods to topological stars and describe their physics. In Sec.~\ref{sec:Leaver} we apply Leaver's method to compute highly damped modes. In Sec.~\ref{sec:conclusion} we summarize our conclusions and possible directions for future work. In Appendices~\ref{app:WKBGen}
and \ref{App:ImaginaryCorrection} we present the derivation of our WKB results, and in Appendix~\ref{app:Leaver} we give details of the calculation based on Leaver's method.

\subsection*{Main Results}

Topological stars are spherically symmetric, charged solutions of five-dimensional Einstein-Maxwell theories, asymptotic to four-dimensional Minkowski plus a compact circle of small size. They are produced by the collapse of the compact circle at a specific locus of the four-dimensional spacetime, forcing the geometry to smoothly end on the surface of a topological bubble. In four dimensions, they appear as singular ultracompact objects that are indistinguishable from charged black holes from afar.  

We study massless scalar perturbations of topological stars. The background is such that the scalar field can be decomposed into a radial wave function that carries the physics of the modes, with the angular dependence decomposed in spherical harmonics with angular momentum $\ell$ and azimuthal index $m$.  For QNMs, the radial wave function must satisfy Neumann boundary conditions at the smooth end of spacetime, and outgoing boundary conditions at spatial infinity. This constrains the mode frequencies to be in a discrete tower labeled by a positive integer $N$, the overtone index.

As shown in Ref.~\cite{Heidmann:2022ehn}, there are two types of topological stars.  The first has a single unstable photon sphere; the second has two photon spheres, an outer (unstable) one and an inner (stable) one.

Here we summarize our results for each of the two types of topological stars.

\vspace{0.2cm}
\paragraph*{\bf Topological stars of the first kind:}

\begin{figure}[t]
\centering
\includegraphics[width=0.5\textwidth]{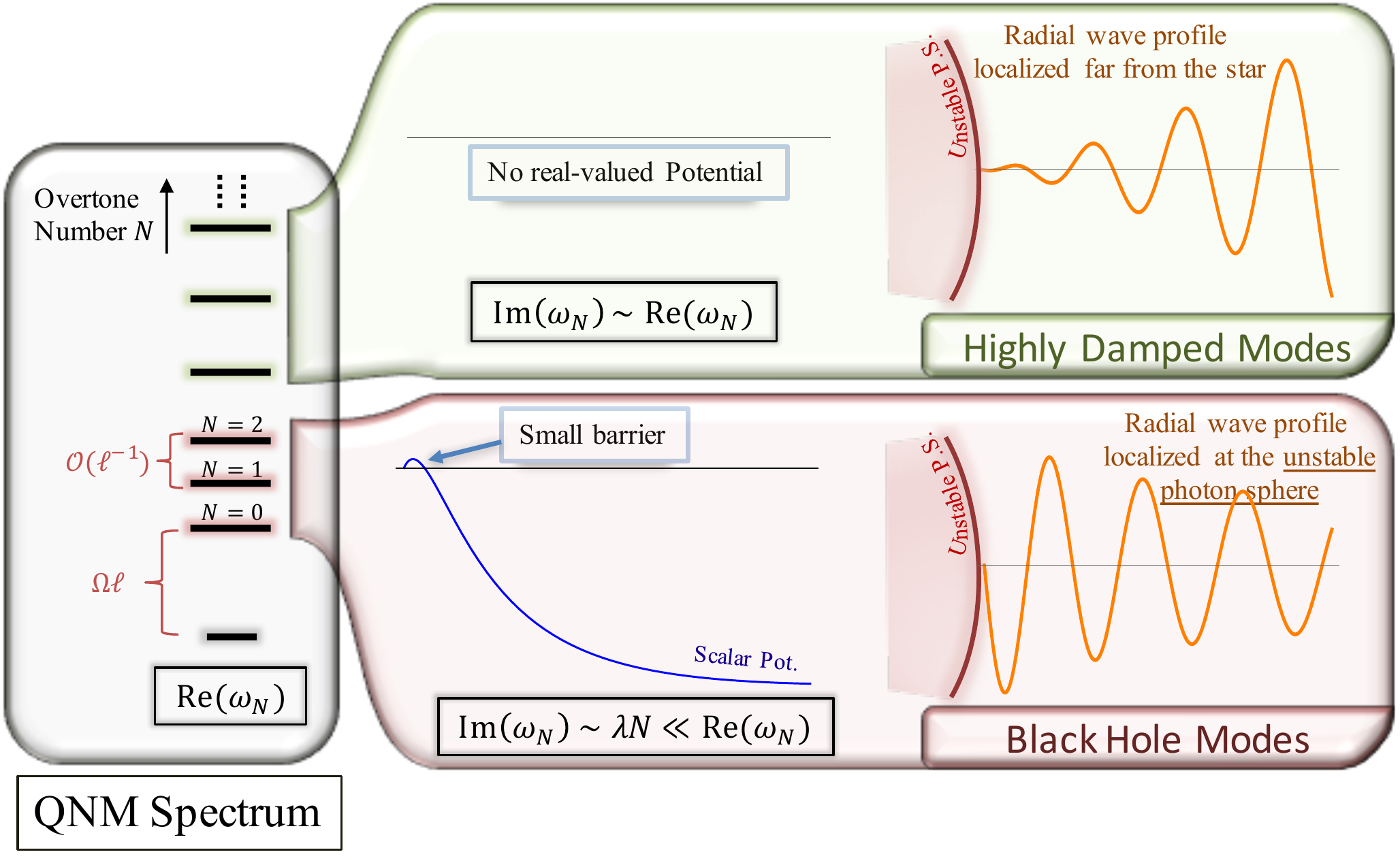}
\caption{QNM spectroscopy of a topological star of the first kind. The left panel describes the mode spectrum and the gaps in the normal frequencies. The two categories of QNMs are described in the right panels, where we illustrate the form of their potential and radial wave profiles.}
\label{fig:IntroTS1}
\end{figure}

Topological stars of the first kind have a single, unstable photon orbit, just like black holes. 
However, the spacetime ends smoothly there and there is no horizon. 
We characterize the spectrum of QNMs by developing a WKB method similar in spirit to the one used for black holes~\cite{Schutz:1985km,Iyer:1986np}, and we also compute it numerically by applying Leaver's method~\cite{Leaver:1985ax}. The spectrum consists of two classes of QNMs, as summarized in Fig.~\ref{fig:IntroTS1}: black hole-like modes and highly damped modes.  

%%%%%%%%%%%%%%%%%%%%%%%%%%%%%%%%
%%%%%%%%%%%%%%%%%%%%%%%%%%%%%%%%

\begin{itemize}
    \item \underline{Black hole-like modes:}
\end{itemize}
    At a given $\ell$, the slowly damped QNMs have an imaginary part much smaller than the real part. As a consequence, the imaginary part of the potential is subleading compared to its real part (see bottom panel in Fig.~\ref{fig:IntroTS1}), and the WKB approximation can be applied. We find that the radial wave profile is localized at the unstable photon orbit, and the spectrum in the eikonal limit, at large and fixed $\ell$, is given by a formula [Eq.~\eqref{eq:IntroEq}] analogous to the formula for black holes (the gaps in the spectrum are also illustrated in Fig.~\ref{fig:IntroTS1}). This spectrum does not depend on the presence or absence of an event horizon, and is purely governed by the unstable photon sphere.
    
\begin{itemize}
    \item \underline{Highly damped modes:}
\end{itemize}

As we go to higher values of $N$,
the modes are not localized anymore (see the top right panel in Fig.~\ref{fig:IntroTS1}). They are highly damped, such that the imaginary part of the frequency cannot be neglected, and we compute them using Leaver's method. This part of the spectrum differs from the black hole spectrum: the real part of the frequency grows with $N$, while the real part of the frequency of highly damped black hole QNMs approaches a constant~\cite{Nollert:1993zz,Motl:2002hd,Motl:2003cd}. However, these modes damp very rapidly and are not expected to contribute significantly to the time-domain signal.

\vspace{0.2cm}
\paragraph*{\bf Topological stars of second kind:}

\begin{figure}[t]
\centering
\includegraphics[width=0.5\textwidth]{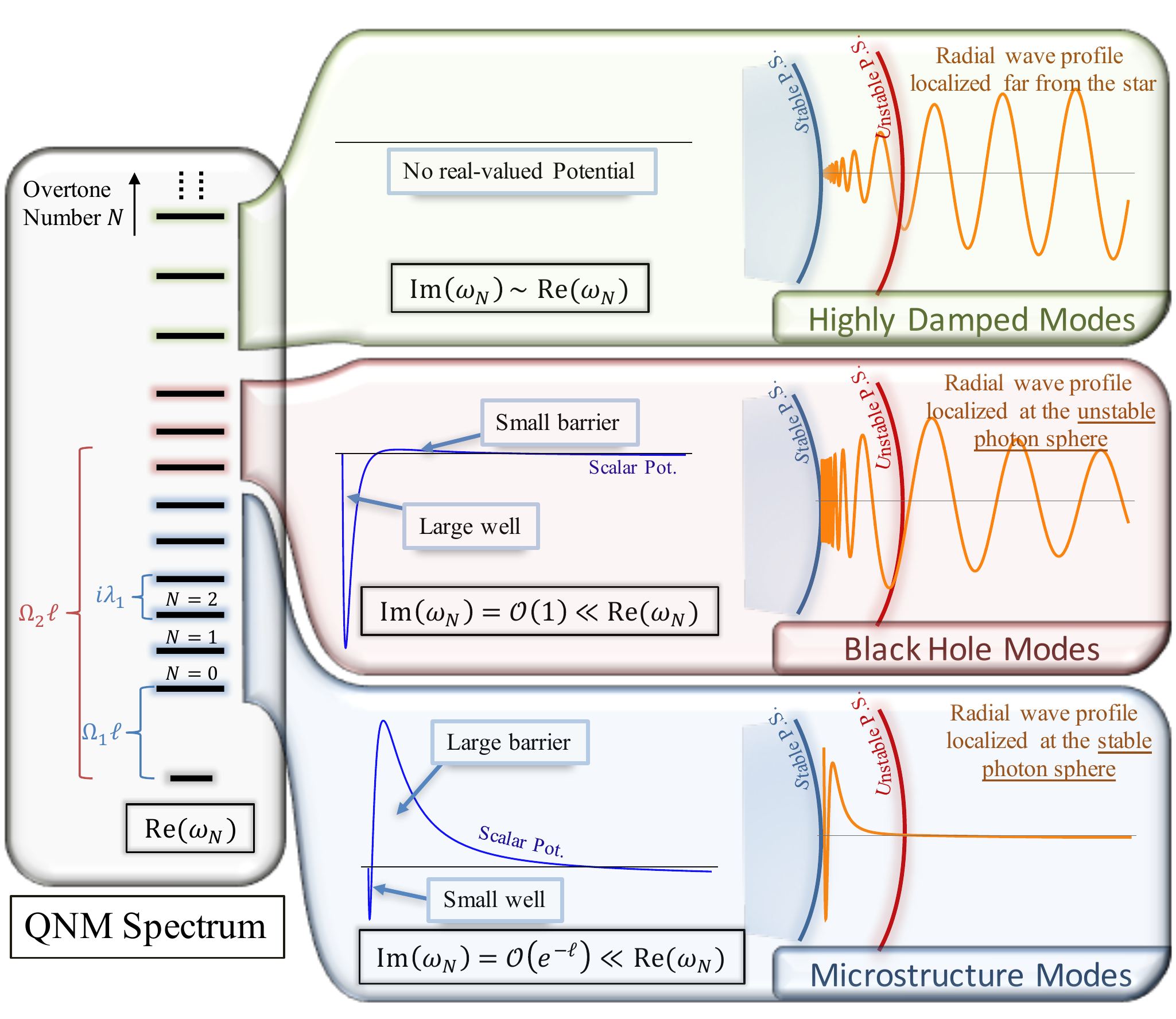}
\caption{QNM spectroscopy of a topological star of the second kind. The overtone number and the different gaps in the normal frequencies are shown on the left panel. The right panels detail the three categories of QNMs.}
\label{fig:IntroTS2}
\end{figure}

Topological stars of the second kind have two photon spheres: the outer photon sphere is unstable, while the inner one is stable.
We develop WKB methods inspired by Refs.~\cite{Bena:2019azk,Bena:2020yii}, and we also apply Leaver's method to compute the QNM spectrum numerically. The QNMs belong to three categories, as summarized in Fig.~\ref{fig:IntroTS2}.  

%%%%%%%%%%%%%%%%%%%%%%%%%%%%%%%% 
%%%%%%%%%%%%%%%%%%%%%%%%%%%%%%%%  
\begin{itemize}
    \item \underline{Microstructure modes:}
\end{itemize}
    The fundamental modes are localized at the inner stable photon orbit and characterized by its scattering properties (see the bottom right panel in Fig.~\ref{fig:IntroTS2}). Since the orbit is stable, they are not only slowly damped, but almost trapped, and their imaginary frequency is exponentially suppressed as a function of $\ell$. In this limit, the real part of the potential is large compared to the imaginary part. The potential has a small well close to the inner photon orbit, and a large barrier prevents the modes from leaking out. Remarkably, the QNM frequencies satisfy a relation similar to Eq.~\eqref{eq:IntroEq} in the eikonal limit:
    \begin{equation}
        \omega_N \,\sim\, \Omega_{1} \,(\ell+\tfrac{1}{2}) - i\,\lambda_1  \,(N+\tfrac{1}{2}) + \frac{\lambda_1}{4\pi} \,e^{-\beta \,\ell},
    \end{equation}
where $\Omega_1$ is the angular velocity of photons at the stable orbits.  The parameter $\lambda_1$ is now purely imaginary ($-i\lambda_1 >0$), and measures the oscillation frequency of geodesics about the stable photon ring. It has the same functional dependence on the geodesic potential $V$ as the Lyapunov exponent of unstable photon orbits, i.e. $\lambda_i=\sqrt{\frac{1}{2 \dot{t}^2} \frac{d^2 V}{d r^2}}$.  
This formula can be regarded as a generalization of the black hole formula to spacetimes with stable photon orbits.    
    
\begin{itemize}
    \item \underline{Black hole modes:}
\end{itemize}
   As we increase $N$, the potential well becomes deeper and deeper while the barrier becomes smaller, and the modes are localized further away from the inner photon sphere. When the barrier is almost vanishing, we will show below that the modes are localized at the outer unstable photon sphere and determined by the scattering properties there (see the middle right panel in Fig.~\ref{fig:IntroTS2}). In the eikonal limit, the frequencies are given by
   \begin{align}
\omega_{N_\text{max}-N} \sim &\,\,\Omega_{2} \,(\ell+\tfrac{1}{2}) +  i\lambda_1 \left( N+\tfrac{1}{2} \right)  \nonumber \\
&+ \frac{\lambda_1}{4\pi} \,\exp \left[\frac{i \pi \lambda_1 (2N+1) }{\lambda_2}\right],
   \end{align}
where $\Omega_2$ and $\lambda_2$ are the angular velocity of photons and their Lyapunov exponent at the outer unstable photon sphere, respectively, and $N_\text{max}$ is the critical overtone number at which the potential barrier disappears. Therefore, the normal frequencies are mainly given by $\Omega_2 \ell$, as they would be for black hole modes with the same unstable photon sphere. Black hole-like modes are therefore contained in the spectrum. However, the imaginary frequencies depend on a nontrivial interplay between both photon spheres, and they are generically smaller than those of black holes. Therefore, having a smooth interior with a stable photon sphere instead of a horizon affects the damping time of the modes, making it generically larger compared to those of black holes. 
\begin{itemize}
    \item \underline{Highly damped modes:}
\end{itemize}
    The upper part of the spectrum corresponds to highly damped modes (see the top right panel in Fig.~\ref{fig:IntroTS2}). They have properties similar to the highly damped modes of topological stars of the first kind. \\

In the conclusions, we will summarize our results and list possible directions for future  research. Moreover, we will discuss the nonlinear instabilities commonly associated with modes localized on stable photon orbits that have recently challenged the relevance of several exotic ultra-compact objects~\cite{Cunha:2022gde,*Keir:2014oka}. We will discuss how the fate of these instabilities may differ in the context of topological solitons, which are coherent quantum gravity states in string theory.

\section{Topological Stars}\label{sec:TSGen} 

Topological stars are solutions of five-dimensional Einstein-Maxwell theory~\cite{Bah:2020ogh,Bah:2020pdz}. This theory arises naturally from the reduction of various supergravity bosonic actions,  so that smooth solutions admit a clear description as coherent states in string theory~\cite{Heidmann:2021cms}. The solutions are asymptotic to $\mathbb{R}^{1,3}\times$S$^1$, so that a coordinate, parametrized by $y$, corresponds to an extra compact circle on top of four-dimensional spacetime. Its asymptotic radius is denoted by $R_{y}$, such that the periodicity of $y$ is given by $y=y+2\pi R_y$.  

Electromagnetic flux is required to support the topology.  We restrict to flux produced by magnetic charges. These can be understood as hidden dark charges that only interact gravitationally.

\begin{figure}[t]
\begin{center}
\includegraphics[width=0.9\columnwidth]{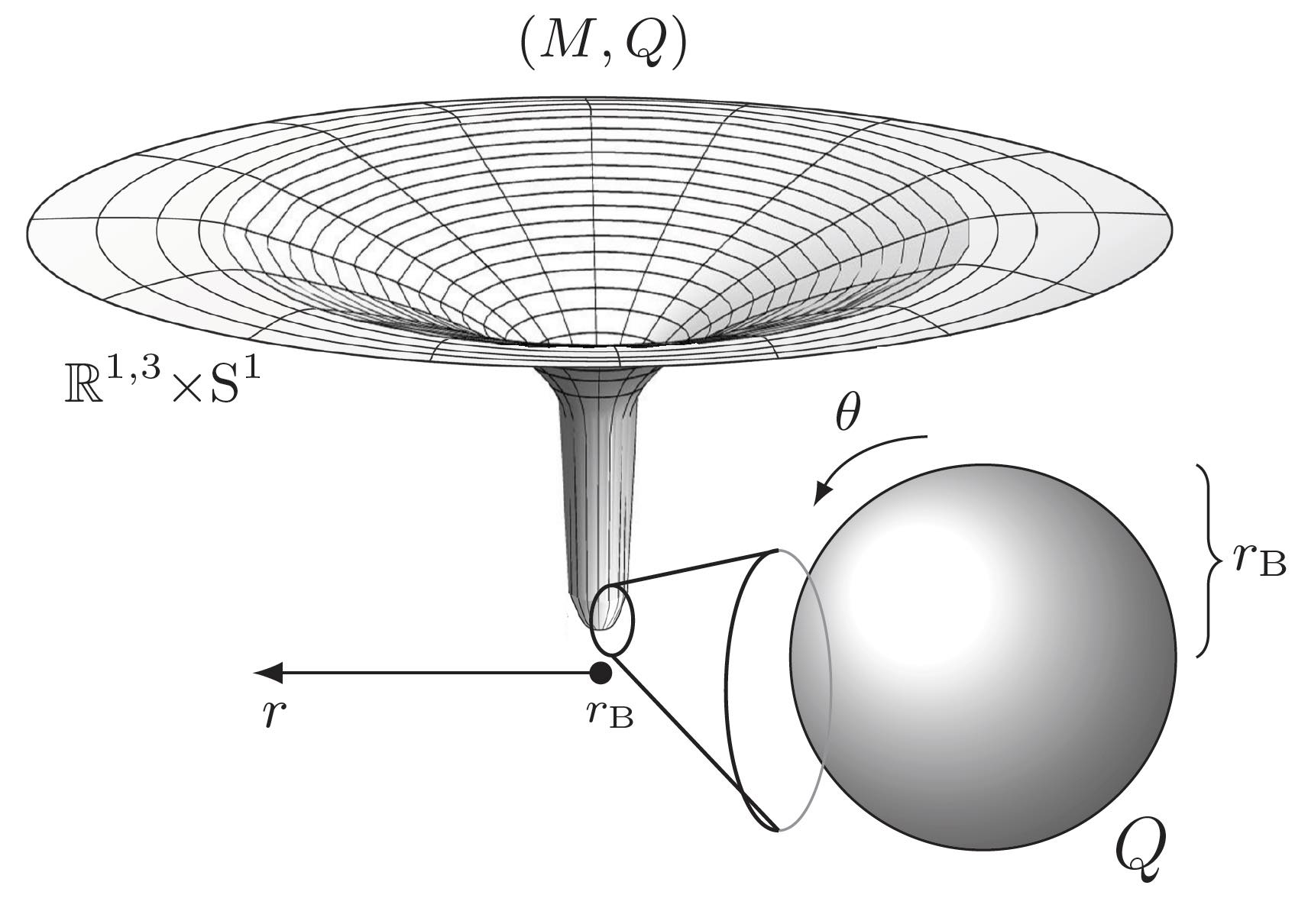}
\caption{Schematic description of a topological star.}
\label{fig:Schematic}
\end{center}
\end{figure}

\subsection{The solution}

Topological stars are static, smooth horizonless geometries that are spherically symmetric and given in terms of two parameters $r_\text{B} > r_\text{S}\geq 0$:
\begin{equation}
\begin{split}
ds_5^2 \,=\, &-\left( 1- \frac{r_\text{S}}{r}\right) \,dt^2+\frac{dr^2}{\left( 1- \frac{r_\text{S}}{r}\right) \left( 1- \frac{r_\text{B}}{r}\right) } \\
&+ r^2 \left(d\theta^2+\sin^2\theta d\phi^2\right) + \left( 1- \frac{r_\text{B}}{r}\right) \,dy^2\,,\\
F \,=\,& \sqrt{3 r_\text{B} r_\text{S} }\,\sin\theta\, d\theta \wedge d\phi\,.
\end{split}
\label{eq:met&GF}
\end{equation}
The spacetime is smooth and terminates at $r=r_\text{B}$ as a coordinate degeneracy of the $y$ circle.  The local topology is best described in terms of a local radial coordinate to the end-to-spacetime locus
\begin{equation}
\rho^2 \equiv \frac{4r_\text{B}^2}{r_\text{B}-r_\text{S}}\, (r-r_\text{B})\,, \qquad \psi\,\equiv \, \frac{\sqrt{r_\text{B}-r_\text{S}}}{2 r_\text{B}^\frac{3}{2}}\,y\,,
\label{eq:localcoor}
\end{equation}
and we consider the limit at $\rho=0$
\begin{equation}
ds_5^2|_{dt=0} \,\sim \,d\rho^2 +\rho^2 \,d\psi^2 + r_\text{B}^2 \left(d\theta^2+\sin^2\theta d\phi^2\right)\,.
\label{eq:localmetric}
\end{equation}
At the end-to-spacetime locus,  the topology is a S$^2$ bubble of radius $r_\text{B}$, defining the geometric size of the topological star.  Figure~\ref{fig:Schematic} is a schematic depiction of the spacetime of a topological star.

Topological stars are inherently four-dimensional gravitational objects that are generated by nontrivial dynamics of an extra compact dimension.  Upon reduction along $y$, they are described by singular geometries with ADM mass $M$ and magnetic charge $Q$ given by (assuming $G_4=1$)
\begin{equation}
M \,=\,  \frac{2r_\text{S} + r_\text{B}}{4}\,,\qquad Q ^2 \,=\, 3  r_\text{B} r_\text{S} \,.
\label{eq:ADMmass}
\end{equation}
The solutions correspond to classically and thermodynamically meta-stable states of gravity if~\cite{Stotyn:2011tv,Bah:2021irr}
\begin{equation}
r_\text{S}\,<\, r_\text{B} \,<\,  2\,r_\text{S}\,.
\label{eq:stabilityrange}
\end{equation}
Moreover,  a topological star approaches its extremal limit when $r_\text{S}\lesssim r_\text{B}$, where it starts to be more and more indistinguishable from the extremal black string,  given by Eq.~\eqref{eq:met&GF}  with $r_\text{S}=r_\text{B}$~\cite{Bah:2020ogh}.

\subsection{Photon spheres}
\label{sec:PhotonSpheres}

\begin{figure}[t]
\begin{center}
\includegraphics[width=0.9\columnwidth]{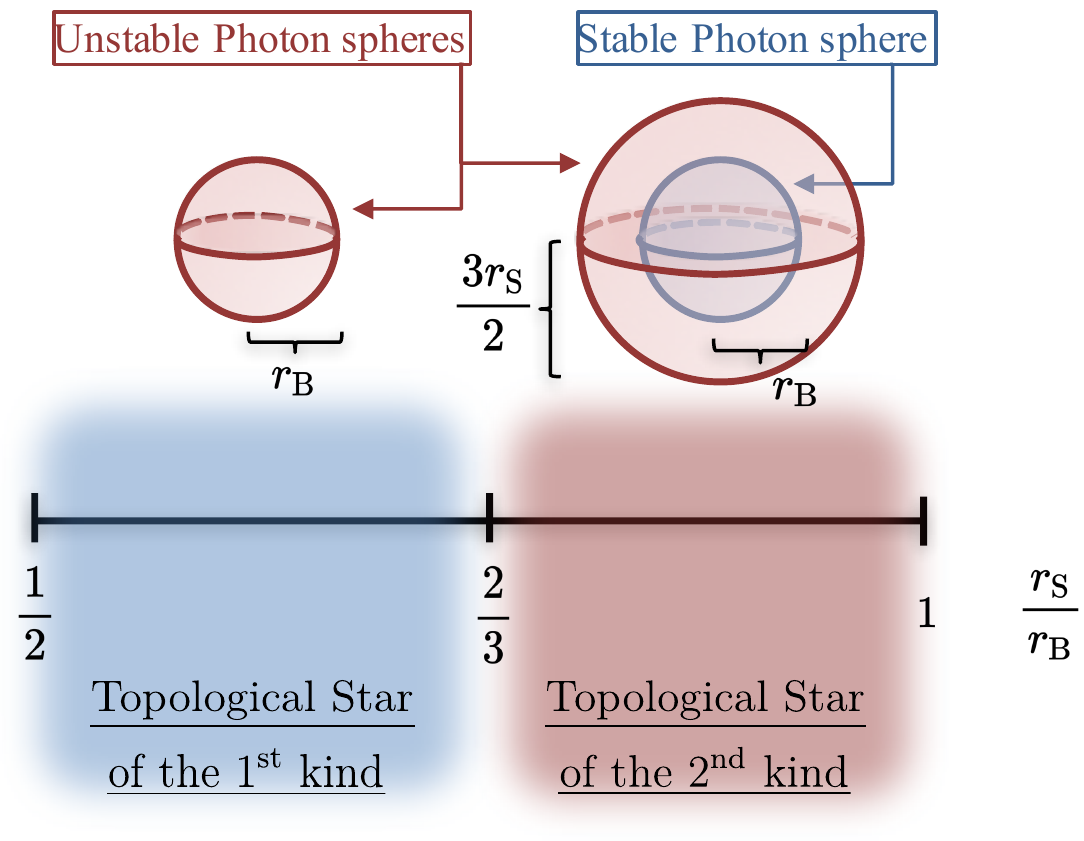}
\caption{Photon spheres of topological stars depending on the range of $r_\text{S}$ and $r_\text{B}$ and charge-to-mass ratio.}
\label{fig:PhotonSpheres}
\end{center}
\end{figure}

The properties of null geodesics scattered by topological stars have been derived in Ref.~\cite{Heidmann:2022ehn}. There are at most two photon spheres, depending on whether $3 r_\text{S} /2$ is greater or smaller than $r_\text{B}$. They are characterized by a radius $R_i$, an angular velocity $\Omega_i$, and a Lyapunov exponent $\lambda_i= \left.\sqrt{\frac{1}{2 \dot{t}^2} \frac{d^2 V_{\mathrm{TS}}}{d r^2}}\right|_{r=R_i}$, where $V_{\mathrm{TS}}$ is the radial potential for null geodesics:
\begin{align}
 & \medmath{ R_1 = r_\text{B},\quad \Omega_1 = \frac{\sqrt{r_\text{B}-r_\text{S}}}{r_\text{B}^{\frac{3}{2}}},\quad \lambda_1 = \frac{\sqrt{\left(r_{\mathrm{B}}-r_{\mathrm{S}}\right)\left(2 r_{\mathrm{B}}-3 r_{\mathrm{S}}\right)}}{r_{\mathrm{B}}^2},}\nonumber \\
&\medmath{ R_2 = \frac{3}{2}\,r_\text{S},\quad \Omega_2 = \frac{2}{3\sqrt{3}\,r_\text{S}},\quad \lambda_2 =\frac{2 \sqrt{3 r_{\mathrm{S}}-2 r_{\mathrm{B}}}}{9 r_{\mathrm{S}}^{\frac{3}{2}}}.} 
\label{eq:PhotonSpheres}
\end{align}
 Note that the Lyapunov exponent of the first photon sphere is either real ($2r_\text{B} > 3 r_\text{S}$) if the photon shell is unstable, or imaginary if it is stable ($2r_\text{B} > 3 r_\text{S}$). 
 
There are therefore two kinds of topological stars with one or two photon spheres, as depicted in Fig.~\ref{fig:PhotonSpheres}:
\begin{itemize}
    \item \underline{Topological star of the first kind:}

When $3r_\text{S}/2 < r_\text{B} < 2 r_\text{S}$, the topological star has a single photon shell localized at the origin of spacetime, $r=r_\text{B}$. The Lyapunov exponent is positive and real, so the photon shell is unstable. 

    \item \underline{Topological star of the second kind:}

      When $r_\text{S} < r_\text{B} < 3r_\text{S}/2 $, the topological star has two photon shells at $r=r_\text{B}$ and $r=3r_\text{S}/2$. The inner one is stable, while the outer one is unstable.

\end{itemize}

%===============================================================
\section{Scalar Perturbations}\label{sec:ScalWave}

We consider a massless minimally coupled scalar that obeys the Klein-Gordon equation in five dimensions:
\begin{equation}
\frac{1}{\sqrt{- \det g}}\,\partial_\mu \left( \sqrt{-\det g}\,g^{\mu \nu} \,\partial_\nu \Phi \right) \,=\,0 \,,
\label{eq:EOMGen}
\end{equation}
where $\Phi$ is the wave function that can depend on all coordinates $(t,r,\theta,\phi,y)$. Topological stars are spherically symmetric with a time and $y$ isometry so  we expand in Fourier modes
\begin{equation}
\Phi_{\ell,m,\omega,p}(t,r,\theta,\phi,y) \,=\,  \frac{K(r)}{r-r_\text{S}} \, Y_{\ell}^{\,m}(\theta,\phi)\,e^{i\left(\omega t + p \frac{y}{R_y}\right)}\,,
\label{eq:Fourier}
\end{equation}
where $\omega$ is the frequency of the perturbation,  $p$ is the quantized momentum along the extra dimension,  $Y_{\ell}^{\,m}$ is the spherical harmonic function of degree $\ell$ and order $m$,  and we have rescaled the radial waveform by $(r-r_\text{S})^{-1}$ for convenience.  Without restriction, we assume that
\begin{equation}
\text{Re}(\omega) \,>\, 0 \,,\qquad \ell\geq 0\,.
\end{equation}

\subsection{Master equation}

The radial wave equation governing $K(r)$,  obtained by inserting Eq.~\eqref{eq:Fourier} into Eq.~\eqref{eq:EOMGen},  is given by
\begin{align}
&\frac{r-r_\text{B}}{r-r_\text{S}}\, \partial_r \left( \frac{r-r_\text{B}}{r-r_\text{S}}\,\partial_r \,K\,\right)  - V(r) \,K \,=\, 0\,, \nonumber\\
&V(r)\equiv \frac{\left(r-r_\text{B}\right)}{\left(r-r_\text{S} \right)^{4}} \left[r_\text{B}-r_\text{S} +\ell(\ell+1) \,(r-r_\text{S})-\omega^2\,r^3\right] \nonumber \\
&\hspace{1.2cm} +\frac{r^3}{\left(r-r_\text{S} \right)^{3}} \,\frac{p^2}{R_y^2}\,, \label{eq:PotGen}
\end{align}
where we recall that $r$ ranges between the smooth origin of spacetime and infinity,  $r_\text{B} \leq r\leq \infty$.  The equation can be written in a Schr\"odinger form, $\partial_{r^*}^2 K- V K=0$,   with the following change of coordinate:
\begin{equation}
r^* \,\equiv\,  r-r_\text{B} + (r_\text{B} -r_\text{S}) \log (r-r_\text{B}).
\end{equation}
The new variable ranges between the origin of spacetime at $r^*=-\infty$ and $r=+\infty$.  

Moreover,  we aim to describe light modes that propagate from the topological star to spatial infinity. This requires $V(+\infty) = p^2/R_y^2-\omega^2 <0$.  From a four-dimensional perspective,  the momentum along the extra dimension acts like an effective mass.  Moreover, a topological star is phenomenologically relevant if the length scale $R_{y}$ is small, say a few orders of magnitude larger than the Planck length. Thus, waves with momentum along the extra dimension have very high energy, of order $p^2/R_{y}^2$, and are very unlikely to be excited in any physical process.  From now on,  we restrict to scalar waves with no momentum along $y$, $p=0$, and the final form of the potential is
 \begin{equation}
 V(r)\equiv \frac{\left(r-r_\text{B}\right)}{\left(r-r_\text{S} \right)^{4}} \left[r_\text{B}-r_\text{S} +\ell(\ell+1) \,(r-r_\text{S})-\omega^2\,r^3\right].
 \label{eq:Pot}
 \end{equation}
\subsection{Boundary conditions for quasi-normal modes}

The QNMs of smooth horizonless geometries correspond to regular waves at the end-to-spacetime locus, and purely outgoing waves at spatial infinity.  

At $r=r_\text{B}$,  the $y$-circle degenerates smoothly as a polar angle degeneracy \eqref{eq:localmetric}.  Regular waves must then satisfy Neumann boundary conditions in terms of the local radius  $\rho$ defined in Eq.~\eqref{eq:localcoor}, $\partial_\rho \Phi \,=\,0\,$, which yields the following condition on the radial wave function $K$:
%$r$ and $r^*$ as
\begin{equation}
\sqrt{r-r_\text{B}} \left[\partial_r K - \frac{K}{r_\text{B}-r_\text{S}} \right] =  e^{-\frac{r^*}{2(r_\text{B}-r_\text{S})}} \partial_{r^*} K =0\,.
\label{eq:RegOrigin}
\end{equation}
Thus, $K$ must be finite as $r\to r_\text{B}$ or $r^*\to -\infty$.\footnote{The second condition requires a priori that $K$ goes to a constant faster than $e^{\frac{r^*}{2(r_\text{B}-r_\text{S})}}$ in terms of $r^*$, but this is guaranteed by the fact that $K(r)$ is finite and differentiable in the limit $r\to r_\text{B}$.}

At a large distance, the potential converges towards $-\omega^2$, and the radial wave function admits two branches proportional to $e^{\pm i\,\omega r^*}$.  Outgoing modes require the ``minus'' branch, such that the scalar field is asymptotically
\begin{equation}
\Phi \underset{r\to\infty}{\sim} \frac{Y_{\ell}^{\,m}(\theta,\phi)}{r}\,e^{i\omega(t-r)}\,.
\label{eq:OutgoingBC}
\end{equation} 

The inner and outer boundary conditions are only satisfied if $\omega$ takes values in a tower of QNM frequencies,  $\omega_N$ and $N\in \mathbb{N}$,  for each value of $\ell$.  The real parts correspond to oscillation frequencies.  The imaginary parts are related to the damping time if $\text{Im}(\omega) >0$ and the mode is stable. If $\text{Im}(\omega) <0$, the mode corresponds to a scalar instability.  Since topological stars are classically stable states in the range considered [see Eq.~\eqref{eq:stabilityrange}],  we do not expect such a mode to exist. 

\section{WKB approximation for slowly damped modes}\label{sec:WKB}

First, we develop WKB methods to calculate QNMs of smooth horizonless geometries for which the wave equation is separable. WKB is an accurate approximation for a Schr\"odinger problem if 
the potential is mostly real-valued,  which requires  $\text{Re}(\omega) \gg \text{Im}(\omega)$,  and if it does not fluctuate wildly.
This is usually achieved in the eikonal limit at large $\ell$ and when the modes are slowly damped.  

Under these assumptions,  the potential can be treated as a real potential for which the physics is dictated by classical turning points,  that are its zeroes.  In between the turning points,  the wave function is given as a superposition of the following waveforms
\begin{equation}
K_\pm (r^*) ~=~ \left|V(r^*)\right|^{-\frac{1}{4}}\, \exp\left[ \pm \int^{r^*}  \sqrt{V(r^*)} \,dr^*\right]\,.
\end{equation}
The different waveforms are connected to each other at the turning points by solving the equation locally and by asymptotic matching.

The topological star potential has turning points given by the roots of the cubic in Eq.~\eqref{eq:Pot} and an asymptotic turning point at the origin, $r^*=-\infty$ ($r=r_\text{B}$).  Because the cubic has no $r^2$ term,  it has a maximum of two real roots in the range $r_\text{B}<r$ depending on the value of $\omega$ and $\ell$. We have therefore developed two WKB methods depending on whether the potential has one or two turning points. The main results are given in the next sections. Details of the derivation can be found in Appendix \ref{app:WKBGen}.

In Sec.~\ref{sec:CondSlowDampTS}, we will see that a topological star of the first kind (one single photon sphere) has a potential with a single root at best, while a topological star of the second kind (two photon spheres) has either two or zero roots.

\subsection{Potential with two roots}
\label{sec:WKB1}

We consider a potential with two roots, $r^*_0$ and $r_1^*$, in between the origin of spacetime and spatial infinity, and we divide it into three zones as depicted in Fig.~\ref{fig:Pot}. The derivation of the WKB spectrum of QNMs is given in Appendix \ref{app:WKB} and is similar in spirit to the technique developed in Refs.~\cite{Bena:2019azk,Bena:2020yii}.  

%%%%%%%%%%%%%%%%%%%%%%%%%%%%%%%%
\begin{figure}[t]
\centering
\includegraphics[scale=0.5]{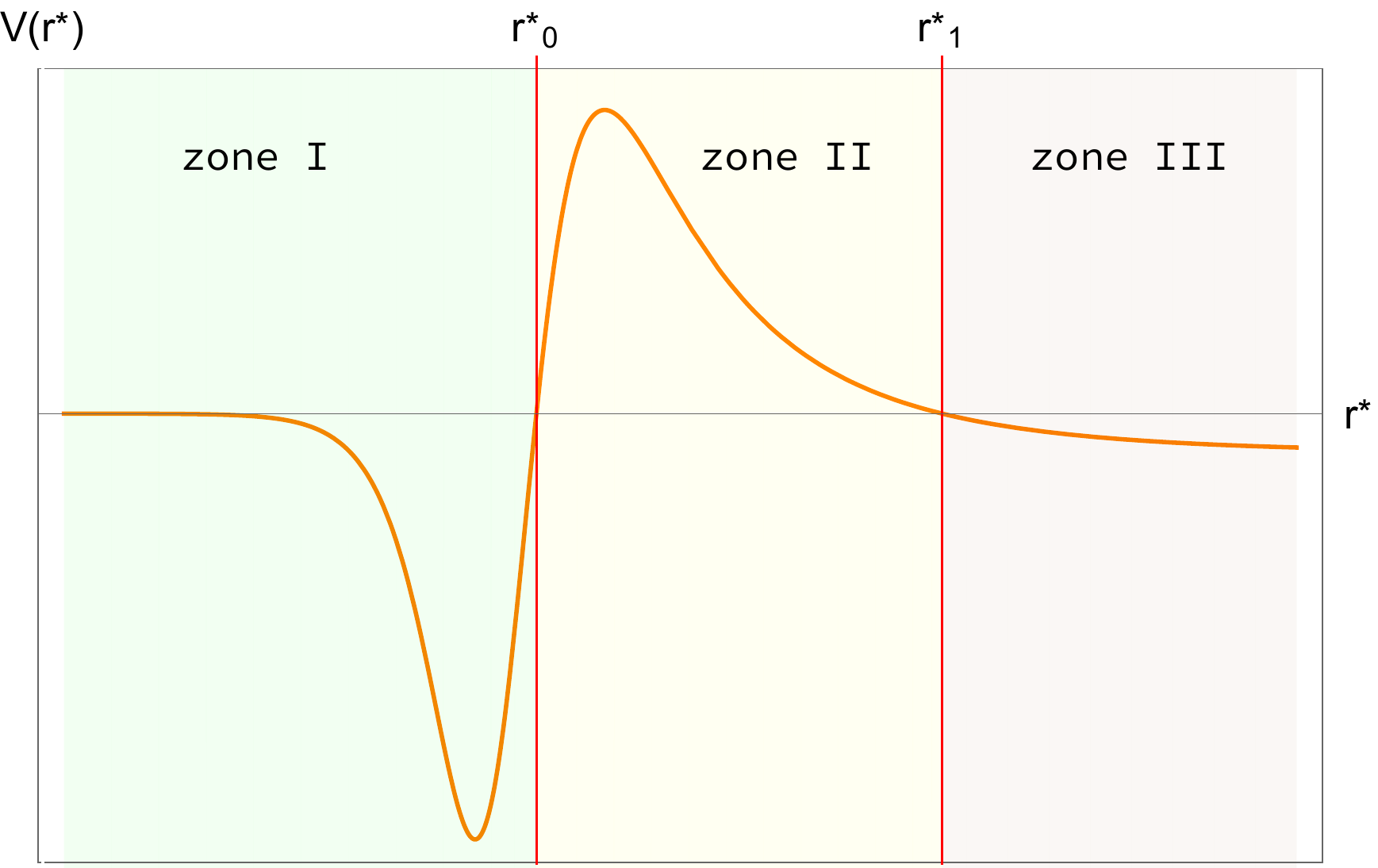}
\caption{Typical form of the potential $V(r^*)$ with two real roots and one asymptotic root at $r^*\to -\infty$.  It has three zones, corresponding to the classical region, the barrier, and the asymptotic region. }
\label{fig:Pot}
\end{figure}
%%%%%%%%%%%%%%%%%%%%%%%%%%%%%%%%

We introduce the well integral,  $\Theta$,  and barrier integral, $T$,  such that
\begin{equation}
\Theta \,\equiv\,  \int_{-\infty}^{r^*_0} |V(r^*)|^{\frac{1}{2}}\,dr^* \,,\qquad T \,\equiv\,  \int_{r_0^*}^{r_1^*} |V(r^*)|^{\frac{1}{2}}\,dr^*\,.
\label{eq:Theta&Tdef1}
\end{equation}
The boundary conditions at the origin and at infinity are compatible when
\begin{equation}
\cos \Theta \,+\, i\, \frac{e^{-2T}}{4}\,\sin \Theta \,=\,0,
\label{eq:QNMWKBGen}
\end{equation}
In the limit of slowly damped modes for which WKB applies, the complex part is a small correction, $e^{-T}\ll 1$.  Thus,  the tower of WKB QNMs is indexed by an integer $N$ such that
\begin{equation}
\omega_N \,=\,  \bar{\omega}_N + \delta \omega_N\,,
\end{equation}
where $\bar{\omega}_N$ are the normal frequencies, and $\delta \omega_N$ are the first-order imaginary corrections determined by
\begin{equation}
\begin{split}
\Theta \,\Bigl|_{\omega= \bar{\omega}_N} &\,=\, \frac{\pi}{2} \,+\,N \pi\,,\\
\delta \omega_N& \,=\, \frac{i}{4} \,\left[\left(\frac{\partial \Theta}{\partial \omega} \right)^{-1} \,e^{-2 T}\right]_{\omega=\bar{\omega}_N}\,.
\end{split}
\label{eq:WKBspectrum}
\end{equation}

\subsection{Potential with a single root}
\label{sec:WKB2}

%%%%%%%%%%%%%%%%%%%%%%%%%%%%%%%%
\begin{figure}[t]
\centering
\includegraphics[scale=0.5]{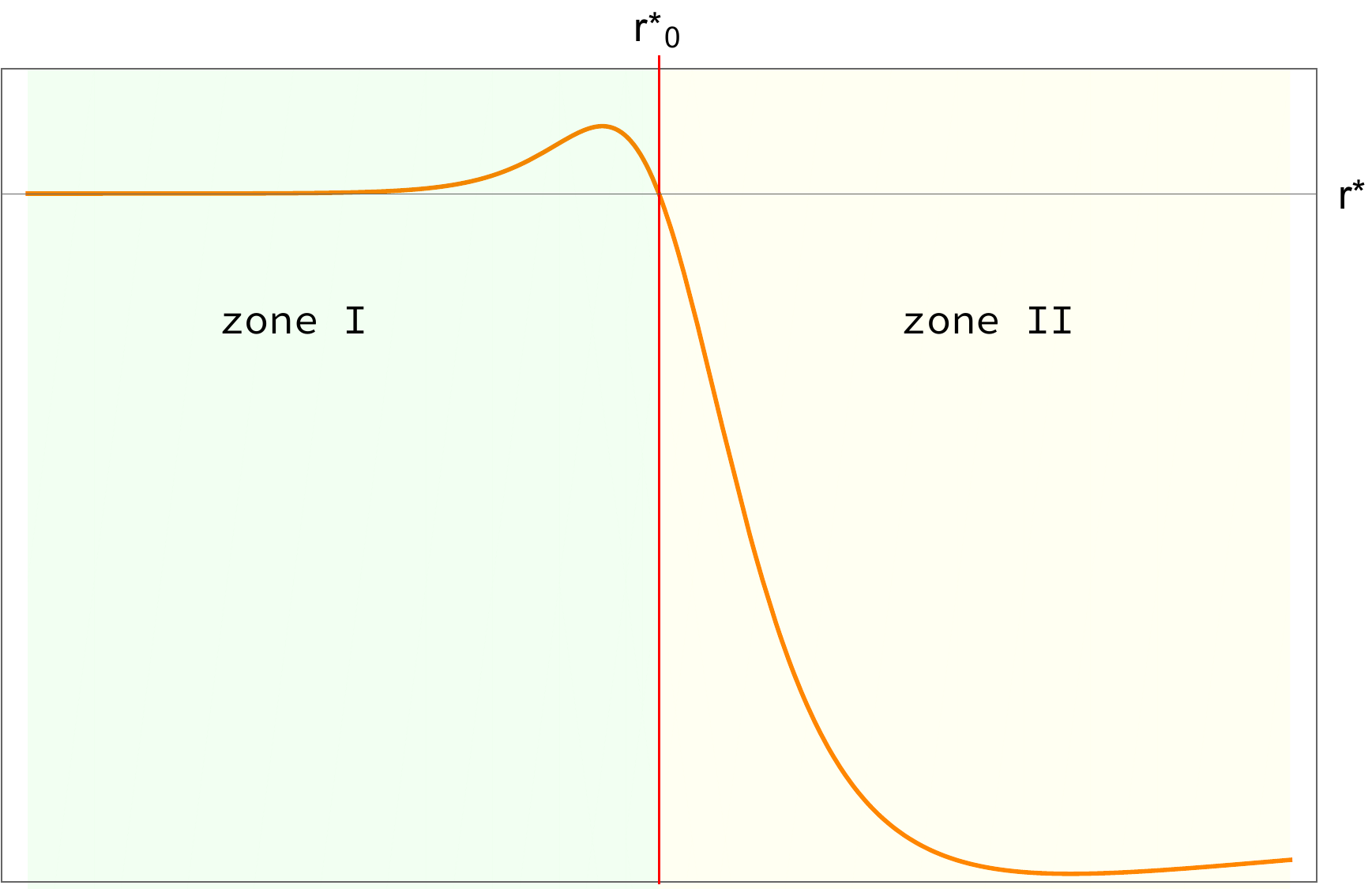}
\caption{Typical form of the potential $V(r^*)$ with a single real root and one asymptotic root at $r^*\to -\infty$.  It has two zones, corresponding to the barrier and the asymptotic region. }
\label{fig:Pot2}
\end{figure}
%%%%%%%%%%%%%%%%%%%%%%%%%%%%%%%%

We now consider a potential with a single root in between the origin of spacetime and spatial infinity. The potential is negative asymptotically and it consists mainly of a potential barrier, as depicted in Fig.~\ref{fig:Pot2}.

At first sight, the WKB approximation cannot capture QNMs. Indeed, the barrier induces a scaling factor of $e^{-T}$ between the transmitted and reflected amplitudes, whereas having purely outgoing modes asymptotically requires both amplitudes to be of the same order. However, this also occurs for black hole QNMs. In Refs.~\cite{Schutz:1985km,Iyer:1986np}, QNMs have been derived when the barrier is small such that $e^{-T}$ is not too large.

In Appendix \ref{app:WKB2}, we develop a WKB calculation in a similar spirit to Refs.~\cite{Schutz:1985km,Iyer:1986np}, but we adapt it to smooth geometries without horizon. We find a tower of modes, labeled by an integer $N$, that is governed by the following algebraic constraint
\begin{equation}
    \quad i  \frac{\sqrt{a}}{2 b \sqrt{c}} \= N + \frac{1}{2} \,, 
   \label{eq:WKBspectrum2}
\end{equation}
where $a,b,c$ are constants determined by the generic expansion of the potential at the origin 
\begin{equation}
V(r^*) \underset{r^*\to -\infty}{\sim} a \,e^{b\,r^*}(1-c \,e^{b\,r^*})\,.
\label{eq:localPot}
\end{equation}
The spectrum of Eq.~\eqref{eq:WKBspectrum2} is valid when the above expansion accurately describes the potential in the whole barrier, i.e. $V(-b^{-1}\log c) \approx 0$. Since the maximum of the potential is well approximated by its asymptotic form, one can relate $a,b,c$ to the properties of the potential there. We find
\begin{equation}
 i \,\frac{\sqrt{2}\, V(r^*_\text{max})}{\sqrt{|V''(r^*_\text{max})|}} \= N + \frac{1}{2} \,, 
\end{equation}
Remarkably, the spectrum is similar to the WKB formula for black holes~\cite{Schutz:1985km,Iyer:1986np}, while the boundary condition applied at the origin of spacetime is very different.

\section{Slowly damped modes of topological stars}
\label{sec:TSQNM}

We apply the generic WKB spectra derived in the previous section to the scalar potential \eqref{eq:Pot} of topological stars. Since we have two different methods depending on the number of zeroes, we first discuss when the conditions necessary to apply the WKB methods are satisfied.

\subsection{Conditions on slowly damped modes}
\label{sec:CondSlowDampTS}

The potential \eqref{eq:Pot} has two roots larger than $r_\text{B}$ if 
\begin{align}
&2 r_\text{B} \left(1+\frac{1}{2(1+\ell(\ell+1))} \right) < 3 r_\text{S}\,, \label{eq:CondonQNMs}\\
& |\omega| < \frac{2(\ell(\ell+1))^\frac{3}{2}}{3\sqrt{3}(r_\text{S}(\ell(\ell+1)+1)-r_\text{B})} \sim \Omega_2 \,(\ell+\tfrac{1}{2})\,, \label{eq:CondonQNMsbis}
\end{align}
where $\Omega_2$ is the angular velocity of photons at the second photon sphere \eqref{eq:PhotonSpheres}.
The first condition implies $$ 2\,r_\text{B} \,<\, 3 \,r_\text{S}.$$ Therefore, the potential can have two roots larger than $r_\text{B}$ only for topological stars of the second kind, as defined in Sec.~\ref{sec:PhotonSpheres}.

This can be interpreted as follows. A potential with a classical interior region, where modes can be localized, separated from the asymptotic region by a large potential barrier, will have modes for which the imaginary part of the frequency is suppressed by a factor $e^{-2T}$. These modes are not only slowly damped, but almost trapped, since their damping time will be very long. In the eikonal limit, these modes are associated with stable photon spheres, where photons can be trapped in a similar way. Since a topological star of the first kind does not have stable orbits, we do not expect such modes to exist. Therefore,  topological stars of the second kind are expected to have fundamental modes localized near the stable photon sphere, that depend on the scattering characteristics of null geodesics there.

The second condition in Eq.~\eqref{eq:CondonQNMsbis} gives an upper bound to the frequencies of slowly damped modes in topological stars of the second kind. Therefore, the number of slowly damped modes is necessarily finite. Note that
some of these modes will have a frequency $|\omega_{N_\text{max}}| \sim \Omega_2 \ell$. Thus, some of the slowly damped modes are connected to the unstable outer photon orbit and to the scattering properties of photons there. This can be understood from the perspective of the potential. For the first modes, the potential has a small well and a large barrier. As we increase the overtone number, the well becomes deeper and the barrier smaller. The upper limit in \eqref{eq:CondonQNMsbis} corresponds to the point at which the barrier disappears.

To summarize, we expect two different categories of slowly damped modes for topological stars of the second kind. The first class of fundamental modes has a large potential barrier inducing a very large damping time. These modes are almost trapped and reflect the internal structure of the background at the inner (stable) photon sphere. The second class of modes has a small  potential barrier and should escape in a much shorter time. Since the form of the potential for these modes resembles the potential of black hole modes~\cite{Schutz:1985km,Iyer:1986np}, we expect them to be black hole-like modes, for which the dynamic is determined by the outer (unstable) photon sphere.

Second, the scalar potential \eqref{eq:Pot} has a single root that is close to $r_\text{B}$ if 
\begin{equation}
    2\, r_\text{B} \,>\, 3\, r_\text{S}\,,\qquad |\omega| < \frac{\sqrt{r_\text{B}-r_\text{S}}}{r_\text{B}^\frac{3}{2}}\,(\ell+1) \sim \Omega_1 \,(\ell+1)\,,
    \label{eq:CondonQNMs2}
\end{equation}
where $\Omega_1$ is the angular velocity of the photon at the inner stable photon sphere \eqref{eq:PhotonSpheres}.
Therefore, one can apply the WKB formula \eqref{eq:WKBspectrum2} to topological stars of the first kind and derive their slowly damped modes. These will still have $\text{Re}(\omega) \gg \text{Im}(\omega)$, but their imaginary part will not be exponentially suppressed by a large potential barrier. As black holes, topological stars of the first kind only have an unstable photon sphere where modes can be localized, but the damping time will not be as extreme as if the orbit were stable. Therefore, we expect these modes to be very similar to the ones derived for black holes. The upper bound shows that their frequencies will be related to scattering properties at the unstable photon orbit.

\subsection{Modes of topological stars of the first kind}
\label{sec:WKBTS1}

We consider the regime of parameters given in Eq.~\eqref{eq:CondonQNMs2}. This corresponds to topological stars of the first kind, and the scalar potential has a small potential barrier, as shown in Fig.~\ref{fig:Pot2}. We apply the generic derivation of Sec.~\ref{sec:WKB2}.

First, we compute the local potential barrier given by three positive constants $(a,b,c)$, as defined in Eq.~\eqref{eq:localPot}. For the topological star potential \eqref{eq:Pot}, we find
\begin{align}
    a &\= \frac{(r_\text{B}-r_\text{S})(\ell(\ell+1)+1)-r_\text{B}^3 \omega^2}{(r_\text{B}-r_\text{S})^4}, \quad b \= \frac{1}{r_\text{B}-r_\text{S}}\,,\nonumber\\
    c &\= \frac{(r_\text{B}-r_\text{S})(3\ell(\ell+1)+4)-r_\text{B}^2(r_\text{B}+3r_\text{S}) \omega^2}{(r_\text{B}-r_\text{S})\left((r_\text{B}-r_\text{S})(\ell(\ell+1)+1)-r_\text{B}^3 \omega^2\right)}.
\end{align}
The WKB spectrum of QNMs is given by Eq.~\eqref{eq:WKBspectrum2}, which we can solve analytically. We find solutions for $0\leq N \leq N_\text{max}$ with
\begin{equation}
    N_\text{max} \sim \frac{\sqrt{3+2\ell(\ell+1)-3 R (1+\ell(\ell+1))}}{1+3 R}-\frac{1}{2},
\end{equation}
where $R=r_\text{S}/r_\text{B}$, and ranges from $\frac{1}{2}$ to $\frac{3}{2}$. The tower of frequencies is given by
\begin{align}
    \omega_N &\= \omega^R_{N} + i \,\omega^I_N \,,\\
    \omega^R_{N} &\= \medmath{\Omega_1\sqrt{N+\tfrac{1}{2}}\,\Biggl[\sqrt{\frac{(\ell(\ell+1)+1)^2}{4(N+\tfrac{1}{2})^2}+4+3\ell(\ell+1)}}\nonumber \\
    & \hspace{2.2cm} +(N+\tfrac{1}{2})(1+3 R)+\frac{\ell(\ell+1)+1}{2(N+\tfrac{1}{2})} \Biggr]^\frac{1}{2} ,\nonumber \\
    \omega^I_{N} &\= \medmath{\Omega_1\sqrt{N+\tfrac{1}{2}}\,\Biggl[\sqrt{\frac{(\ell(\ell+1)+1)^2}{4(N+\tfrac{1}{2})^2}+4+3\ell(\ell+1)}}\nonumber \\
    & \hspace{2.2cm} -(N+\tfrac{1}{2})(1+3 R)-\frac{\ell(\ell+1)+1}{2(N+\tfrac{1}{2})} \Biggr]^\frac{1}{2} .\nonumber
\end{align}

In the eikonal limit, we retrieve the same relation as for black hole QNMs:
\begin{equation}
\omega_N \underset{\ell \to \infty}{\sim} \,\Omega_{1} \,(\ell+\tfrac{1}{2}) + i\,\lambda_1  \,(N+\tfrac{1}{2})\,,
\label{eq:WKBQNMTS1}
\end{equation}
where $\Omega_1$ and $\lambda_1$ are the angular velocity and Lyapunov exponent of the photon orbit at $r=r_\text{B}$ \eqref{eq:PhotonSpheres}. A topological star of the first type has a single photon sphere and is unstable, so the term proportional to the Lyapunov exponent gives an imaginary contribution, as for black holes. Therefore, the spectrum of modes is identical to that of a black hole with the same photon sphere while having different internal boundary conditions. This is quite a remarkable result, considering that a topological star is smooth and horizonless, without any absorption: it shows that the fundamental modes that govern the response of black holes under perturbations are independent of having a horizon and are mainly given by the scattering properties of its photon ring.

\subsection{Modes of topological stars of the second kind}

We now consider the regime of parameters given by Eq.~\eqref{eq:CondonQNMs} and \eqref{eq:CondonQNMsbis}. The generic derivation of Sec.~\ref{sec:WKB1} can be applied to topological stars of the second kind. We first derive the normal frequencies, and then compute the imaginary corrections of Eq.~\eqref{eq:WKBspectrum}.

\subsubsection{Normal frequencies}

The well integral \eqref{eq:Theta&Tdef1}  gives
 \begin{align}
\Theta &= \int_{r_\text{B} }^{r_0} \frac{1}{r-r_\text{S}} \sqrt{\frac{\omega^2\, r^3 -\ell(\ell+1) \,(r-r_\text{S})+r_\text{S}-r_\text{B}}{r-r_\text{B}}} \,dr\,, \nonumber
\end{align}
where $r_0$ is the first of the two positive real roots of the cubic polynomial. Unfortunately,  cubic roots have a rather complicated expression, and the integral can be expressed only in terms of elliptic functions.  However,  we can perform an approximation. In the range \eqref{eq:CondonQNMs},  the cubic is mainly a linear function in between $r_\text{B}$ and $r_0$ such that 
\begin{equation}
\Theta \sim \pi \left(\sqrt{\frac{r_\text{B}^2(3 r_\text{S}-2 r_\text{B})}{r_\text{B}-r_\text{S}} \omega^2 - 1} - \sqrt{\ell(\ell+1)-3 r_\text{B}^2 \omega^2} \right). \nonumber
\end{equation}
The normal frequencies are therefore given by Eq.~\eqref{eq:WKBspectrum}:
\begin{align}
\bar{\omega}_N = &\medmath{\Omega_1\,\sqrt{2N+1}  \Biggl[(N+\tfrac{1}{2})\left(3 R-\tfrac{5}{2}\right)+ \frac{ 1+\ell(\ell+1) }{2N+1}} \label{eq:NormalFreq}\\
&\hspace{0.1cm} \medmath{ + \sqrt{(3 R-2 ) \left(\ell(\ell+1)-3(N+\tfrac{1}{2})^2(1-R)\right)-3(1-R)}\Biggr]^{\tfrac{1}{2}} },  \nonumber
\end{align}
where $\Omega_1$ is the angular velocity of trapped photons at the inner photon sphere \eqref{eq:PhotonSpheres}, and we used again $R=r_\text{S}/r_\text{B}$ with now $2/3 < R <1$. The upper bound fixed by Eq.~\eqref{eq:CondonQNMs} gives the total number of modes at a given $\ell$:
\begin{equation}
\label{eq:max_modes_cond}
\frac{2N_\text{max} +1}{\sqrt{\ell(\ell+1)}} \,\sim\, \frac{2}{3R} \sqrt{\frac{R-\tfrac{2}{3}}{1-R}}\left(2-\sqrt{3(3 R+2)(1-R))} \right),
\end{equation}
  The number of slowly damped modes is therefore increasing as the topological star approaches its extremal limit, $R\to 1$. 
This is a consequence of having ultra-compact, near-extremal smooth geometries that are almost indistinguishable from an extremal black string. The topological stars develop a very long black hole throat that caps off smoothly, where many modes can pile up. 

Moreover, far from extremality, the normal frequencies scale like ultra-compact objects of mass $M$, as given by Eq.~\eqref{eq:ADMmass}:
\begin{equation}
\bar{\omega}_N \= \mathcal{O}\left(\frac{(\ell,N)}{M} \right)  \,,
\end{equation}
while a change of scales occurs near extremality:
\begin{equation}
\bar{\omega}_N \propto \frac{\sqrt{\epsilon}\,(N+\ell)}{M} \,,
\end{equation}
where $\epsilon$ is the nonextremality parameter $Q=\frac{4M}{\sqrt{3}}(1+\epsilon)$, with $\epsilon\ll 1$.  These very small frequencies mean that the scattered waves will ``echo'' off the star at very long time scales of order $M/\sqrt{\epsilon}$ after having explored the long throat of the geometry \cite{Bena:2019azk,Bena:2020yii}.

The WKB approximation is particularly accurate in the eikonal limit, $\ell \gg 1$. We first consider the eikonal limit for the first few fundamental modes, where $N$ is fixed and small. We find, up to order $\ell^0$,
\begin{equation}
   \bar{\omega}_N \underset{\ell \to \infty}{\sim} \,\Omega_{1} \,(\ell+\tfrac{1}{2}) - i\,\lambda_1  \,(N+\tfrac{1}{2}) \,,
\end{equation}
where $\Omega_1$ and $\lambda_1$ are the angular velocity and ``Lyapunov'' exponent of the stable photon orbit at $r=r_\text{B}$: see Eq.~\eqref{eq:PhotonSpheres}. We remind that $\lambda_1$ is an imaginary number for a topological star of the second kind so that $-i\lambda_1 >0$.

Now, we consider the eikonal limit for the last modes in the tower. We find, up to order $\ell^0$,
\begin{align}
    \bar{\omega}_{N_\text{max}-N} &\underset{\ell \to \infty}{\sim} \,\Omega_{2} \,(\ell+\tfrac{1}{2}) +  \frac{i\,\lambda_1 }{\mathcal{R}}\,\left( N+\tfrac{1}{2} \right), \nonumber \\
    \mathcal{R}&\equi 2 \sqrt{\frac{1-R}{\tfrac{2}{3}+R}}+3 R-2\,,
    \label{eq:NormalFreqLast}
\end{align}
where $\Omega_2$ is the angular velocity of photons at the outer (unstable) photon sphere: see Eq.~\eqref{eq:PhotonSpheres}. The function $\mathcal{R}$ varies between $1$ and $1.2$ for $2/3<R<1$. This is probably an artifact of our linear approximation of the cubic in the potential, and at this level of approximation, it could make sense to just assume that $\mathcal{R} \approx 1$.

These formulas confirm the qualitative arguments in the previous section. One has a first class of modes (the first fundamental ones) that are localized at the inner stable photon sphere and characterized by the scattering properties there. Moreover, there is a second class of modes (the last slowly damped modes) that explore the geometry up to the outer unstable photon sphere, and are governed by the dynamics of photons there.

\subsubsection{Imaginary corrections}

We now estimate the imaginary corrections to the real (normal) frequencies given by Eq.~\eqref{eq:WKBspectrum}.  One needs to derive the barrier integral $T$ of Eq.~\eqref{eq:Theta&Tdef1}.  Unfortunately, the integral can only be expressed in terms of elliptic functions. Its generic expression is not particularly interesting and is given in Appendix \ref{App:ImaginaryCorrection}. Moreover, we find
\begin{align}
\frac{\partial \Theta}{\partial \omega}\bigg|_{\omega = \bar{\omega}_N} \= \frac{\pi r_\text{B}^2 \bar{\omega}_N}{r_\text{B}-r_\text{S}} &\Biggl[\frac{3(r_\text{B}-r_\text{S})}{\sqrt{\ell(\ell+1)-3r_\text{B}^2 {\bar{\omega}_N}^2}} \\
&+ \frac{3 r_\text{S}-2 r_\text{B}}{\sqrt{\ell(\ell+1)-3r_\text{B}^2 {\bar{\omega}_N}^2}+N+\tfrac{1}{2}} \Biggr] \,.\nonumber
\end{align}

First,  we consider the eikonal limit for the first class of fundamental modes, $\ell \gg 1$ and $N\ll \ell$.  We find that
\begin{equation}
T \sim  \beta \,\ell\,,\quad 
\frac{\partial \Theta}{\partial \omega}\bigg|_{\omega = \bar{\omega}_N} \sim \frac{\pi\,r_\text{B}^2}{\sqrt{(r_\text{B}-r_\text{S})(3r_\text{S}-2r_\text{B})}}=\frac{-i\pi}{\lambda_1},
\end{equation}
for some constant $\beta$, that depends nontrivially on $r_\text{S}$ and $r_\text{B}$. This constant cannot be related to any geodesic quantities. Indeed, it indicates how slowly the modes leak out and depends on the geometry outside the inner photon sphere.  Finally, we have
\begin{equation}
\delta \omega_N \underset{\ell \to \infty}{\sim}  \frac{\lambda_1}{4\pi} \,e^{-\beta \,\ell}\,.
\label{eq:eikonalImaginary}
\end{equation}
These exponentially suppressed imaginary frequencies are not specific to topological stars, and have been also observed for AdS black holes~\cite{Holzegel:2013kna}, ultra-compact neutron stars~\cite{Keir:2014oka}, various ECOs~\cite{Cardoso:2014sna}, and supersymmetric microstate geometries in string theory~\cite{Eperon:2016cdd,Bena:2020yii}. They are characteristic of modes localized at a stable photon orbit with extreme damping time. By estimating the energy decay of such modes following the references above, we similarly find that they decay like $(\log t)^{-2}$, while modes localized at unstable photon orbits (characteristic of black holes) decay much faster, as $t^{-2}$ \cite{Eperon:2016cdd,Bena:2020yii}.

For the eikonal limit of the second class of modes, one can approximate the potential in the barrier by a quadratic polynomial, $V\sim V(r^*_\text{max}) + \tfrac{1}{2} V''(r^*_\text{max}) (r^*-r^*_\text{max})^2$, and estimate the barrier integral $T$ using this approximation, which gives
\begin{equation}
    T \,\sim \, \frac{\pi \, V(r^*_\text{max})}{\sqrt{2 |V''(r^*_\text{max})|}}\,.
\end{equation}
By considering the large $\ell$ limit, we find
\begin{equation}
    \delta \omega_{N_\text{max}-N} \underset{\ell \to \infty}{\sim}  \frac{\lambda_1}{4\pi\,\mathcal{R}} \,\exp \left[-\frac{(2N+1) \pi}{\mathcal{R}}\,\frac{i\lambda_1}{\lambda_2}\right],
\end{equation}
where $\mathcal{R} \approx 1$ has been defined in Eq.~\eqref{eq:NormalFreqLast}. As expected, these modes are not exponentially suppressed as a function of $\ell$, and therefore are less trapped than the first few fundamental modes. However, one still has $|\delta \omega_{N_\text{max}-N} | \ll \bar{\omega}_{N_\text{max}-N}$, so that the modes are slowly damped (i.e., their damping time is large compared to their period). 

Their energy decay will be of order $t^{-2}$ like black hole modes. However, unlike black holes, the imaginary part results from a nontrivial interplay between the stable and  unstable photon spheres. It is rather intuitive that the damping time depends on the stability of both orbits, which therefore produces a cavity effect. As the inner sphere becomes more stable ($|\lambda_1|$ increases), the imaginary part decreases and the modes have a longer lifetime, while as the outer sphere becomes more unstable ($\lambda_2$ increases), the imaginary part increases and the modes decay more rapidly.

\subsubsection{Eikonal limit}

%%%%%%%%%%%%%%%%%%%%%%%%%%%%%%%%
\begin{figure}[t]
\centering
\includegraphics[scale=0.6]{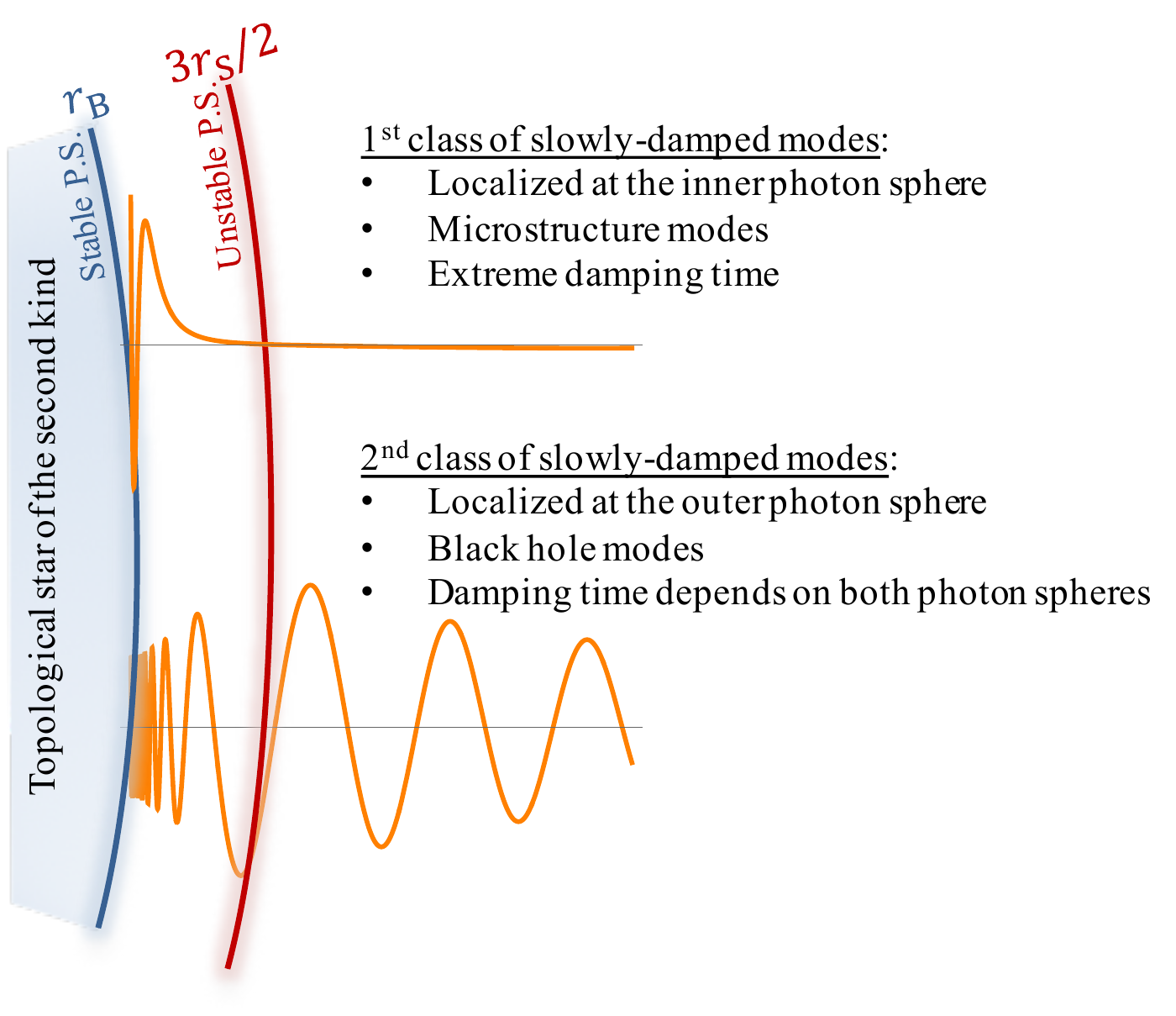}
\caption{Wave profiles of the first and second class of slowly damped modes with $\ell=5$ for a topological star of the second kind.}
\label{fig:QNMTS2}
\end{figure}
%%%%%%%%%%%%%%%%%%%%%%%%%%%%%%%%

The QNM frequencies of black holes are related to geodesics at their photon orbit, i.e., their ``shadow''~\cite{Cardoso:2008bp}. In the eikonal limit, these frequencies are 
\begin{equation}
    \omega_N^\text{BH} \,\sim\, \Omega \, (\ell+\tfrac{1}{2}) + i \,\lambda (N+\tfrac{1}{2})\,,
    \label{eq:BHQNMs}
\end{equation}
where $\Omega$ and $\lambda$ are the angular velocity and Lyapunov exponent of photons at the unstable photon sphere. 

Topological stars of the second kind are smooth horizonless geometries with a stable inner photon sphere and an unstable outer photon sphere. We have found that their QNMs have a richer structure. There are two classes of fundamental modes, given in the eikonal limit by (we consider $\mathcal{R}\approx 1$) 
\begin{align}
    \omega_N &\,\sim\, \Omega_{1} \,(\ell+\tfrac{1}{2}) - i\,\lambda_1  \,(N+\tfrac{1}{2}) + \frac{\lambda_1}{4\pi} \,e^{-\beta \,\ell}\,,\label{eq:FreqEikonalfirst}\\
     \omega_{N_\text{max}-N} &\,\sim \,\Omega_{2} \,(\ell+\tfrac{1}{2}) +  i\,\lambda_1 \,\left( N+\tfrac{1}{2} \right) \label{eq:FreqEikonallast} \\
     &\hspace{0.5cm} + \frac{\lambda_1}{4\pi} \,\exp \left[-\frac{i \pi \lambda_1 (2N+1) }{\lambda_2}\right], \nonumber
\end{align}
where $\Omega_1$ and $\lambda_1$ are the angular velocity and Lyapunov exponent of the stable photon orbit at $r=r_\text{B}$, and $\Omega_2$ and $\lambda_2$ are the angular velocity and Lyapunov exponent of the unstable photon orbit at $r=3r_\text{S}/2$: see Eq.~\eqref{eq:PhotonSpheres}.

The first class of fundamental modes is remarkably similar to the black hole QNMs~\eqref{eq:BHQNMs}, but the identification is at the level of the stable photon sphere, so the ``$-i\lambda$'' term is now a real and positive contribution. The real part of the QNM frequencies at the inner photon sphere is much smaller than the real part of the QNMs at the outer (unstable) photon sphere, $\Omega_1 \ell < \Omega_2 \ell$. These modes carry information about the microstructure of the solutions at the core of the geometry~\cite{Bena:2019azk,Bena:2020yii}.
Their imaginary part is exponentially suppressed as a function of $\ell$, so that they have a very long lifetime.

The second class of slowly damped modes is localized at the unstable photon orbit. The real parts of their QNM frequencies are proportional to $\Omega_2\ell$. These modes can be thought of as ``black hole-like'' modes. Indeed, the modes of a black hole with the same unstable photon sphere will have the same normal frequencies. However, their imaginary part, i.e. their damping time, will be slightly different. This is because the damping time is influenced by what is inside the outer photon sphere, and whether it is an absorbing horizon or a stable photon sphere affects the dynamics. 

Thus, we found that black hole modes in  smooth topological geometries with inner stable photon orbits last longer than one would expect for black holes. This longer damping time in the wave signal produced by a cavity effect in between photon spheres could provide new avenues of exploration for future experiments, different from the echoes highlighted in previous work~\cite{Cardoso:2016rao,Cardoso:2016oxy,Cardoso:2019rvt}.

In Figure \ref{fig:QNMTS2}, we schematically summarize the dynamics of the slowly damped modes of topological stars of the second kind. The first class of fundamental, almost-trapped modes is localized and determined by the inner (stable) orbit; the second class of ``black hole-like'' modes are instead localized around the outer (unstable) orbit. 

From a broader perspective, the WKB spectrum [Eqs.~\eqref{eq:FreqEikonalfirst} and \eqref{eq:FreqEikonallast}] should be understood as the generalization of the black hole result [Eq.~\eqref{eq:BHQNMs}] for any smooth horizonless geometries with stable internal photon orbits surrounded by unstable outer photon orbits.

\subsection{Concluding remarks}

Despite the fact that topological stars are smooth and horizonless, such that everything gets reflected at linear order, the slowly damped QNM frequencies are similar to those of black holes, and they are determined by the scattering properties of null geodesics in the vicinity of photon rings. 

For topological stars of the first kind, which have a single unstable photon orbit and no stable photon orbit, the modes are identical to those of a black hole with the same photon sphere, and given by Eq.~\eqref{eq:WKBQNMTS1}. This has the surprising implication that the fundamental modes of a black hole do not depend on the presence of a horizon, which is rather counterintuitive, considering that the horizon is crucial to determine the boundary conditions for the perturbations.

Topological stars of the second kind have a stable photon sphere surrounded by an unstable one. The first class of modes is determined by the stable orbit: see Eq.~\eqref{eq:FreqEikonalfirst}. The real part of the QNM frequency is small and proportional to $\Omega_1 \ell$~\cite{Bena:2019azk,Bena:2020yii}.
They carry information about the inner, smooth structure of the topological star. Moreover, they are almost trapped at the stable orbit, with a very long damping time. The second class of slowly damped modes is black hole-like: the real part of their QNM frequency is determined by the scattering of photons at the unstable photon sphere, and proportional to $\Omega_2 \ell$. However, their damping time depends on a nontrivial interplay between the inner and outer photon sphere, while it depends only on the Lyapunov exponent of the shadow for a black hole. These modes have a longer damping time than the modes of a black hole. If topological stars have astrophysical relevance, their QNM frequencies may provide new smoking guns to investigate quantum gravity with future experiments.

\section{Leaver's method}\label{sec:Leaver}

The WKB method is a good approximation for QNMs that have a relatively large $\ell$ and $\text{Im}(\omega)\ll \text{Re}(\omega)$. Thus, it fails to capture highly damped modes with $\text{Im}(\omega)\gtrsim \text{Re}(\omega)$,  and the approximation may introduce errors in the spectrum for small $\ell$'s. In this section, we use Leaver's method to numerically compute the QNMs, including the highly damped ones, and we assess the accuracy of the  WKB method when it is applied beyond its regime of validity. We refer the reader to Appendix~\ref{appendix:numerical_methodology} for more details on the method.

\subsection{Leaver's method}

\begin{table*}[t]
\setlength{\tabcolsep}{0.5em}
\begin{tabular}{| l l| c | c | c | c | c |}
\hline \hline  
\makecell{$N$}& \makecell{~} & \makecell{$\ell=0$}& \makecell{$\ell=1$} & \makecell{$\ell=2$} & \makecell{$\ell=3$} & \makecell{$\ell=4$}  \\ \hline \hline
\multirow{2}{*}{$0$}  & {\footnotesize Leaver:} & ${\medmath{0.2398+0.0039i}}$ & ${\medmath{\mathbf{0.4966+(3.2\times10^{-6})i}}}$ & ${\medmath{\mathbf{0.7497+(9.4\times 10^{-10})i}}}$ & ${\medmath{\mathbf{1.0023+(2.2\times 10^{-13})i}}}$ & ${\medmath{\mathbf{1.2547+(4.4\times10^{-17})i}}}$ \\
                     & {\footnotesize WKB:} & -- & ${\medmath{\mathbf{0.5359+(1.2\times10^{-5})i}}}$ & ${\medmath{\mathbf{0.7782+(2.3\times 10^{-9})i}}}$ & ${\medmath{\mathbf{1.0244+(4.2\times 10^{-13})i}}}$ & ${\medmath{\mathbf{1.2727+(7.9\times 10^{-17})i}}}$ \\ \hline
\multirow{2}{*}{$1$}  & {\footnotesize Leaver:} & ${\medmath{0.4477+0.0400i}}$ & ${\medmath{\mathbf{0.7183+0.0006i}}}$ & ${\medmath{\mathbf{0.9804+(5.7\times 10^{-7})i}}}$ & ${\medmath{\mathbf{1.2361+(2.8\times 10^{-10})i}}}$ & ${\medmath{\mathbf{1.4901+(7.0\times10^{-14})i}}}$ \\
                       &{\footnotesize WKB:} & -- & ${\medmath{\mathbf{0.7393+0.0021i}}}$ & ${\medmath{\mathbf{0.9992+(1.3\times10^{-6})i}}}$ & ${\medmath{\mathbf{1.2523+(5.2\times 10^{-10})i}}}$ & ${\medmath{\mathbf{1.5041+(1.7\times 10^{-13})i}}}$ \\ \hline
\multirow{2}{*}{$2$}  & {\footnotesize Leaver:} & ${\medmath{0.6576+0.1081i}}$ & ${*~\medmath{0.9112+0.0150i}~*}$ & ${\medmath{\mathbf{1.1986+(7.9\times10^{-5})i}}}$ & ${\medmath{\mathbf{1.4621+(8.1\times 10^{-8})i}}}$ & ${\medmath{\mathbf{1.7198+(4.8\times10^{-11})i}}}$ \\
                       & {\footnotesize WKB:} & -- & ${*~\medmath{0.9334+0.0935i}~*}$ & ${\medmath{\mathbf{1.2146+(1.8\times 10^{-4})i}}}$ & ${\medmath{\mathbf{1.4756+(1.5\times10^{-7})i}}}$ & ${\medmath{\mathbf{1.7318+(8.3\times10^{-11})i}}}$ \\ \hline
\multirow{2}{*}{$3$}  & {\footnotesize Leaver:} & ${\medmath{0.8223+0.2057i}}$ & ${\medmath{1.0990+0.0699i}}$ & ${*~\medmath{1.3965+0.0036i}~*}$ & ${\medmath{\mathbf{1.6787+(1.0\times10^{-5})i}}}$ & ${\medmath{\mathbf{1.9430+(1.1\times10^{-8})i}}}$ \\
                       & {\footnotesize WKB:} & -- & -- & ${*~\medmath{1.4215+0.0121i}~*}$ & ${\medmath{\mathbf{1.6933+(1.9\times10^{-5})i}}}$ & ${\medmath{\mathbf{1.9550+(1.8\times10^{-8})i}}}$ \\ \hline
\multirow{2}{*}{$4$}  & {\footnotesize Leaver:} & ${\medmath{0.9667+0.1892i}}$ & ${\medmath{1.3138+0.1372i}}$ & ${\medmath{1.5728+0.0335i}}$ & ${\medmath{\mathbf{1.8813+0.0006i}}}$ & ${\medmath{\mathbf{2.1586+(1.3\times10^{-6})i}}}$ \\
                       & {\footnotesize WKB:} & -- & -- & -- & ${\medmath{\mathbf{1.9043+0.0015i}}}$ & ${\medmath{\mathbf{2.1733+(2.3\times 10^{-6})i}}}$ \\ \hline
\multirow{2}{*}{$5$}  & {\footnotesize Leaver:} & ${\medmath{1.1842+0.1959i}}$ & ${\medmath{1.5282+0.1860i}}$ & ${\medmath{1.7599+0.0931i}}$ & ${*~\medmath{2.0611+0.0120i}~*}$ & ${\medmath{\mathbf{2.3638+(8.6\times10^{-5})i}}}$ \\
                       & {\footnotesize WKB:} & -- & -- & -- & ${*~\medmath{2.1075+0.0764i}~*}$ & ${\medmath{\mathbf{2.3862+(1.9\times10^{-4})i}}}$ \\ \hline 
\multirow{2}{*}{$6$}  & {\footnotesize Leaver:} & ${\medmath{1.3950+0.2166i}}$ & ${\medmath{1.7241+0.2231i}}$ & ${\medmath{1.9576+0.1587i}}$ & ${\medmath{2.2334+0.0562i}}$ & ${*~\medmath{2.5520+0.0030i}~*}$ \\
                       & {\footnotesize WKB:} & -- & -- & -- & -- & ${*~\medmath{2.5931+0.0113i}~*}$ \\ \hline
\multirow{2}{*}{$7$}  & {\footnotesize Leaver:} & ${\medmath{1.6034+0.2352i}}$ & ${\medmath{1.9170+0.2438i}}$ & ${\medmath{2.1398+0.2274i}}$ & ${\medmath{2.4205+0.1191i}}$ & ${*~\medmath{2.7202+0.0275i}~*}$ \\
                       & {\footnotesize WKB:} & -- & -- & -- & -- & ${*~\medmath{2.7935+0.1479i}~*}$ \\ \hline
\multirow{2}{*}{$8$}  & {\footnotesize Leaver:} & ${\medmath{1.8144+0.2505i}}$ & ${\medmath{2.1197+0.2573i}}$ & ${\medmath{2.2935+0.2574i}}$ & ${\medmath{2.6169+0.1941i}}$ & ${\medmath{2.8951+0.0811i}}$ \\
                       & {\footnotesize WKB:} & -- & -- & -- & -- & -- \\ \hline 
\multirow{2}{*}{$9$}  & {\footnotesize Leaver:} & ${\medmath{2.0290+0.2644i}}$ & ${\medmath{2.3289+0.2710i}}$ & ${\medmath{2.4835+0.2600i}}$ & ${\medmath{2.8752+0.2565i}}$ & ${\medmath{3.0843+0.1457i}}$ \\
                       & {\footnotesize WKB:} & -- & -- & -- & -- & -- \\ \hline
\end{tabular}

\caption{The ten first QNM frequencies, $\omega_N$, of the topological star of the second kind ($r_\text{B} = 0.68$ and $r_\text{S}=0.66$) for $\ell$ from 0 to 4. The top and bottom lines in each cell are the Leaver and WKB values respectively. We have denoted the separate types of modes in the following way: microstructure modes are bolded, black hole modes are starred, and the highly damped modes are in plain text.
}
\label{tab:QNMs_and_WKB}
\end{table*}

Leaver's method involves a series solution to the radial perturbation equation which may then be used to compute QNM spectra~\cite{Leaver:1985ax}. Starting with the radial perturbation equation \eqref{eq:PotGen} with no momentum ($p=0$), we expand the radial function $K(r)$ in a series: 
\begin{equation}
\label{eq:Leaver_series}
K(r) = e^{-i \omega r} \sum_{n=0}^\infty a_n \left(\frac{r-r_\text{B}}{r-r_\text{S}}\right)^n.
\end{equation}
The solution satisfies the boundary conditions \eqref{eq:RegOrigin} and \eqref{eq:OutgoingBC} when the coefficients $a_n$ are such that the series is convergent. The radial equation translates into a four-term recurrence relation for the $a_n$ coefficients:
 \begin{equation}
 \label{eq:leaver_coeffs}
 \begin{aligned}
 \alpha_n a_{n+1} + \beta_n a_n+\gamma_n a_{n-1}+\delta_n a_{n-2} = 0 ,
 \end{aligned}
 \end{equation}
 where we set $a_{-2}=a_{-1}=0$. Here $\alpha_n$, $\beta_n$, $\gamma_n$ and $\delta_n$ are functions of $r_\text{S}, r_\text{B}, n,\ell$ and $ \omega$, given in Appendix~\ref{appendix:numerical_methodology}. We can perform a Gaussian elimination step to find a three-term recurrence relation with new coefficients:
 \begin{equation}
 \label{eq:leaver_coeffs1}
 \begin{aligned}
 \alpha'_n a_{n+1} + \beta'_n a_n+\gamma'_n a_{n-1} = 0 .
 \end{aligned}
 \end{equation}
The series of $a_n$ can be shown to converge by calculating the ratio $a_{n+1}/a_n$ at large $n$,
\begin{equation}
\label{eq:leaver_convergence}
\frac{a_{n+1}}{a_n} \,\underset{n\to\infty}{\rightarrow}\, 1 \pm \frac{\sqrt{2 i \omega(r_\text{B} - r_\text{S})}}{n^{1/2}}+\frac{ (5r_\text{B}-2r_\text{S})i-3 \omega}{4n}. \nonumber
\end{equation}
 The series converges uniformly when we select the branch corresponding to the minus sign in the equation above. This corresponds to ``minimal solutions,'' and then $\omega$ corresponds to a QNM frequency~\cite{doi:10.1137/1009002, pincherle1892generation, pincherle1894delle}. From Eq.~\eqref{eq:leaver_coeffs1}, the QNM frequencies must satisfy the following infinite continued fraction equation:
\begin{equation}
\label{eq:ContFrac}
\beta'_0 - \frac{\alpha'_{0} \gamma'_1}{\beta'_{1}-}\frac{\alpha'_{1} \gamma'_{2}}{\beta'_{2}-}...\= 0,
\end{equation}
where we have used common notation for continued fractions~\cite{doi:10.1137/1009002,Leaver:1985ax,*Leaver:1990zz} . The equation can be inverted an arbitrary number of times to obtain equivalent conditions. Empirically, the $N^{\text{th}}$ QNM frequency is a stable root of the $N^{\text{th}}$ inversion of the continued fraction~\cite{Leaver:1985ax,*Leaver:1990zz}, defined as
\begin{equation}
\label{eq:Leaver_inversions}
\begin{aligned}
&\left( \beta'_{N} - \frac{\alpha'_{N-1}\gamma'_{N}}{ \beta'_{N-1}-}  \frac{\alpha'_{N-2}\gamma'_{N-1} }{\beta'_{N-2}-} ... \frac{\alpha'_{0}\gamma'_{1}}{\beta'_{0}}\right) \\ 
&\hspace{1cm}- \left(\frac{\alpha'_{N} \gamma'_{N+1}}{\beta'_{N+1}-} \frac{\alpha'_{N+1} \gamma'_{N+2}}{\beta'_{N+2}-}...\right)  = 0,
\end{aligned}
\end{equation}
These algebraic relations can be solved numerically to obtain the QNM spectrum of topological stars.

\subsection{Results and comparison with WKB}

We apply Leaver's method to compute QNM frequencies for the dominant multipoles ($\ell=0,\dots,4$). We consider two illustrative topological stars with mass $M=1/2$, as defined in Eq.~\eqref{eq:ADMmass}:
\begin{itemize}
    \item A topological star of the second kind with $r_\text{B}=0.68$ and $r_\text{S}=0.66$, that is $Q=1.16$ from Eq.~\eqref{eq:ADMmass}. We have deliberately chosen $r_\text{S}\sim r_\text{B}$ so that the topological star is near extremal, with a long capped throat that can support a large number of slowly damped modes, as predicted by the WKB analysis.
    \item A topological star of the first kind with $r_\text{B}=13/14\simeq 0.93$ and $r_\text{S}=15/28\simeq 0.54$, so that $Q\simeq 1.22$.
\end{itemize}

\subsubsection{Topological star of the second kind}
\label{sec:TS2Leaver}

\begin{figure}[ht]
\centering
\includegraphics[width=0.45\textwidth]{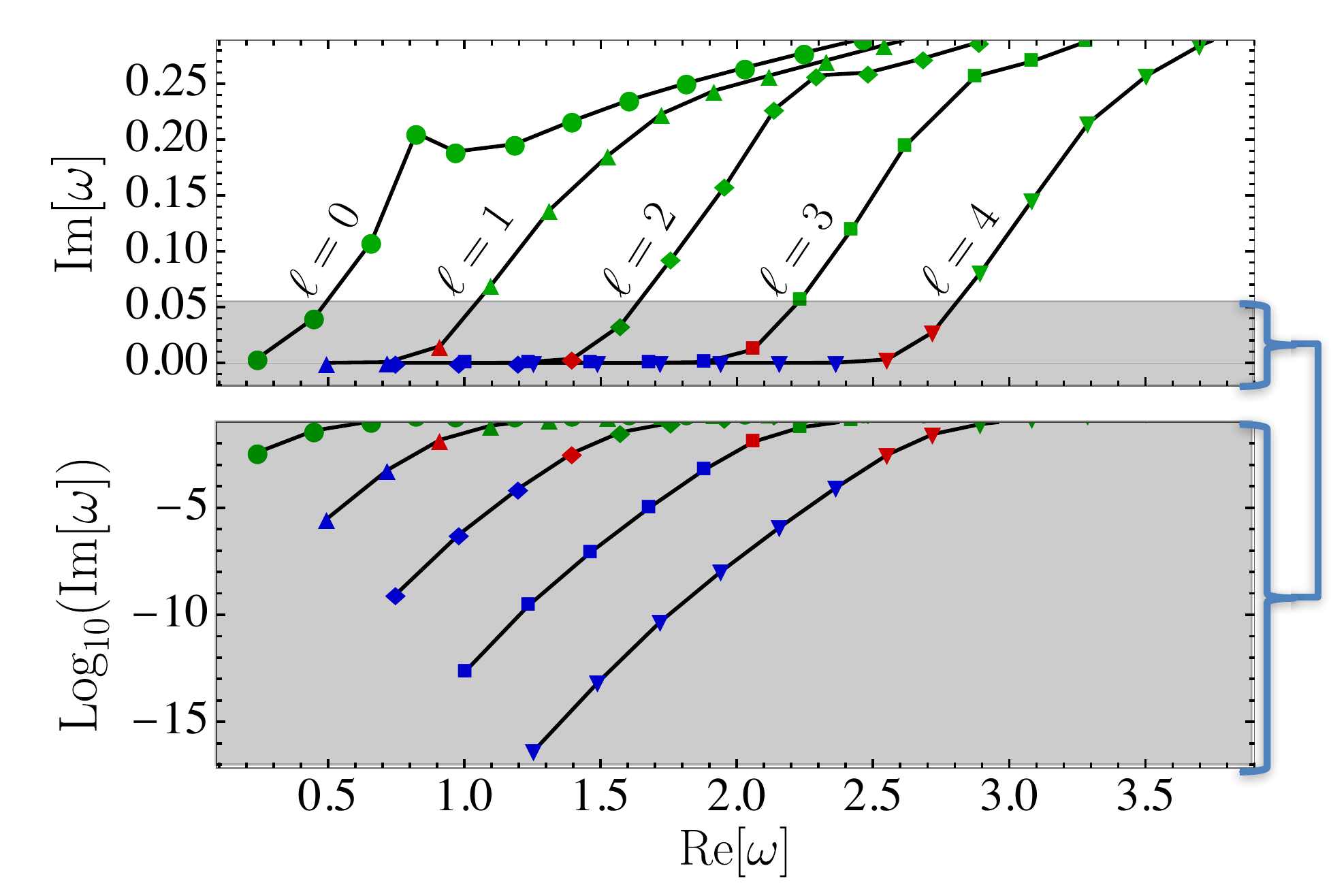}
\caption{The first few QNMs of the topological star of the second kind, with values listed in Table~\ref{tab:QNMs_and_WKB}, with units normalized to $M=1/2$. The modes associated with the inner stable photon sphere are shown in blue. The red points correspond to the black hole modes, and the modes in green correspond to modes inaccessible with our WKB formula, found with Leaver's method.}
\label{fig:2nd_kind_QNMs}
\end{figure}

\begin{figure}[ht]
\centering
\includegraphics[width=0.45\textwidth]{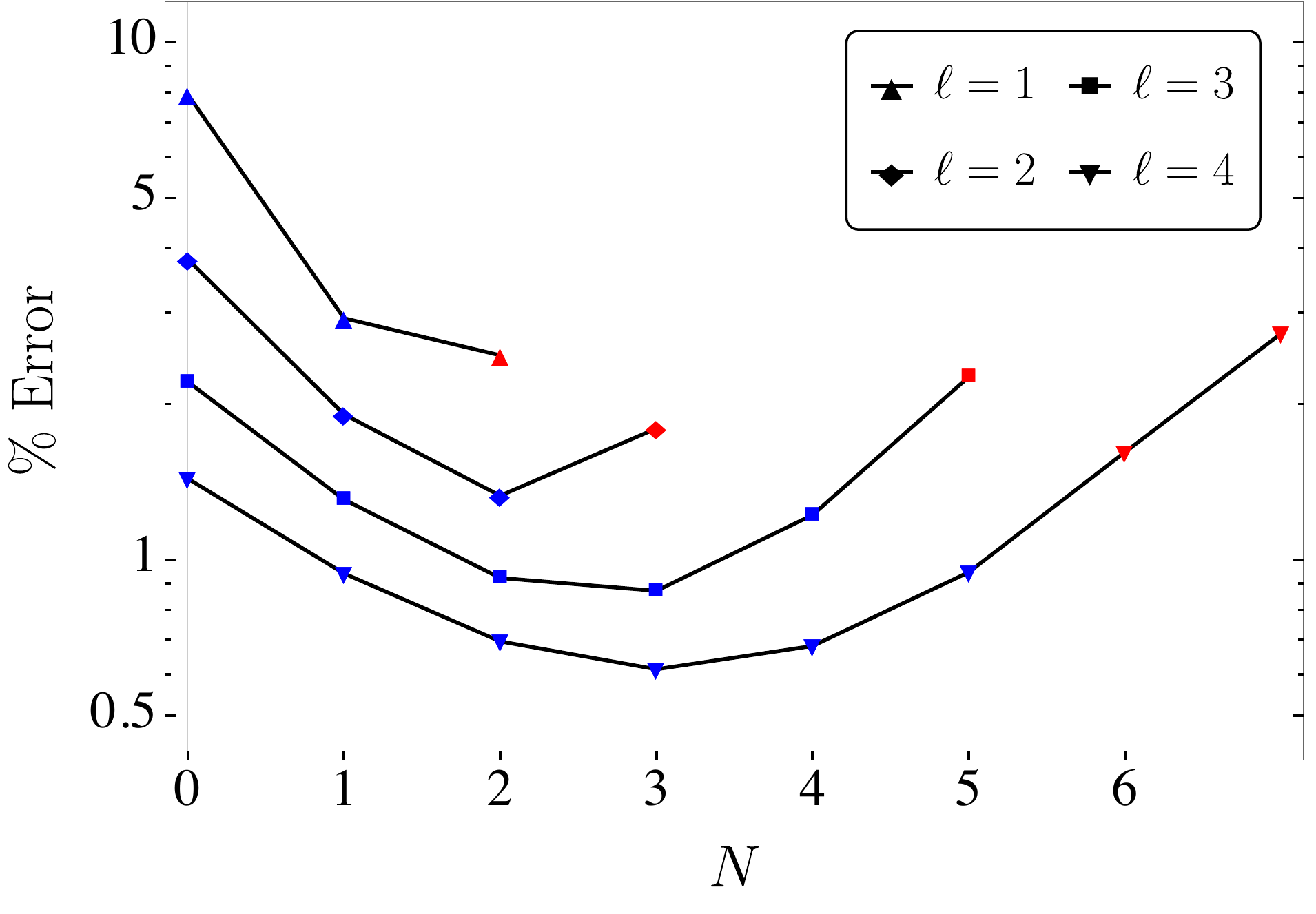}
\caption{Percent error, $\left| \frac{\omega_\text{Leaver}-\omega_\text{WKB}}{\omega_\text{Leaver}} \right|$, between the WKB and Leaver frequencies of Table~\ref{tab:QNMs_and_WKB}, for the topological star of the second kind.}
\label{fig:Percent_Error}
\end{figure}

We first consider the topological star of the second kind given above. The QNM frequencies found by Leaver's method are plotted in Fig.~\ref{fig:2nd_kind_QNMs} for $\ell=0,\dots,4$.
In Table \ref{tab:QNMs_and_WKB} we list the  first ten QNMs, along with their WKB counterparts when they exist.
In Fig.~\ref{fig:Percent_Error} we also plot the relative difference between the Leaver calculation and the WKB approximation. In the figures and table, we use three different point styles or fonts to highlight the mode classification discussed in our WKB calculation:
\begin{itemize}
    \item \underline{Microstructure modes}  (bold and blue):
\end{itemize}    
The first few fundamental modes are localized at the inner stable photon orbit. Their frequencies are determined by the scattering properties there: see Eq.~\eqref{eq:FreqEikonalfirst}.
\begin{itemize}
    \item \underline{Black hole modes} (star and red):
\end{itemize}  
These are the black hole-like modes. They are localized at the unstable outer photon orbit and determined by the scattering properties there: see Eq.~\eqref{eq:FreqEikonallast}.
\begin{itemize}
    \item \underline{Highly damped modes} (plain and green):
\end{itemize}  
As we increase the overtone number $N$, we find an infinite tower of highly damped modes for which the imaginary part of the frequencies cannot be neglected anymore. These modes (listed in plain text in Table~\ref{tab:QNMs_and_WKB}) could not be found by the WKB method for this reason.      
This part of the spectrum differs from the usual highly damped black hole QNMs. For Schwarzschild black holes, when the imaginary part of the QNM frequency (and $N$) increases, the real part of the frequency decreases~\cite{Leaver:1985ax,*Leaver:1990zz}. A similar behavior is observed in Sec.~\ref{sec:BHcomp} below, where we compute the QNMs for a black string with the same mass and similar charge as the topological star. For topological stars, both the real and imaginary parts increase with $N$. This difference in the spectrum should not significantly affect the time-domain signal, since these modes damp out very quickly.

%%%%%%%%%%%%%%%%%%%%%%%%%%%%%%%% 
%%%%%%%%%%%%%%%%%%%%%%%%%%%%%%%%

%%%%%%%%%%%%%%%%%%%%%%%%%%%%%%%%
%%%%%%%%%%%%%%%%%%%%%%%%%%%%%%%%

Figure~\ref{fig:Percent_Error} shows the percentage error between the Leaver and WKB frequencies. The agreement is better as $\ell$ increases, and has a parabola shape as a function of $N$. The latter is due to the fact that, as $N$ approaches $N_\text{max}$ or $0$, the classical turning points are close to each other so that the WKB approximation loses accuracy. Overall, our WKB method is very accurate in capturing mode frequencies even for small $\ell$, so that the error is of the order of 7\% in the worst case, $\ell=1$ and $N=0$, and is already less than 1\% for several $N$ at $\ell=4$.

WKB approximations are known to be relatively inaccurate in the context of black hole modes. However, this is due not so much to the WKB approximation itself,
but rather to the asymptotic matching method applied in that context, which requires matching the inner and outer waveforms by an approximately quadratic potential. The method used here for horizonless solutions, described in Appendix~\ref{app:WKB}, is different in spirit, does not rely on this approximation, and is, therefore, more accurate.

\subsubsection{Topological star of the first kind}

\begin{table*}[ht]
\setlength{\tabcolsep}{0.4em}
\begin{tabular}{| l l | c | c | c | c | c | c|}
\hline \hline  
\makecell{$N$}& & \makecell{$\ell=0$}& \makecell{$\ell=1$} & \makecell{$\ell=2$} & \makecell{$\ell=3$} & \makecell{$\ell=4$} & \makecell{$\ell=10$} \\ \hline \hline
\multirow{2}{*}{$0$} & {\footnotesize Leaver:} & ${\medmath{0.4994+0.2775i}}$ & ${\medmath{1.1290+0.2267i}}$ & ${*~\medmath{1.8088+0.2100i}~*}$ & ${*~\medmath{2.4982+0.2016i}~*}$ & ${*~\medmath{3.1913+0.1966i}~*}$ & ${*~\medmath{7.3753+0.1863i}~*}$\\
                       & {\footnotesize WKB:} & -- & -- & ${*~\medmath{2.0297+0.1221i}~*}$ & ${*~\medmath{2.6592+0.1498i}~*}$ & ${*~\medmath{3.3167+0.1619i}~*}$ & ${*~\medmath{7.4274+0.1779i}~*}$ \\ \hline
\multirow{2}{*}{$1$} & {\footnotesize Leaver:} & ${\medmath{0.8365+0.9839i}}$ & ${\medmath{1.3227+0.8228i}}$ & ${\medmath{1.9656+0.7351i}}$ & ${\medmath{2.6361+0.6879i}}$ & ${\medmath{3.3159+0.6583i}}$ & ${*~\medmath{7.4562+0.5894i}~*}$\\
                       & {\footnotesize WKB:} & -- & -- & -- & -- & -- & ${*~\medmath{7.7859+0.3554i}~*}$ \\ \hline
\multirow{2}{*}{$2$} & {\footnotesize Leaver:} & ${\medmath{0.8935+1.2605i}}$ & ${\medmath{1.6057+1.5284i}}$ & ${\medmath{2.1781+1.3722i}}$ & ${\medmath{2.8192+1.2704i}}$ & ${\medmath{3.4828+1.2038i}}$ &${\medmath{7.5774+1.0402i}}$\\
                       & {\footnotesize WKB:} & -- & -- & -- & -- & -- & -- \\ \hline
\multirow{2}{*}{$3$} & {\footnotesize Leaver:} & ${\medmath{1.1539+1.4666i}}$ & ${\medmath{1.6328+1.8153i}}$ & ${\medmath{2.4483+2.0676i}}$ & ${\medmath{3.0390+1.9215i}}$ & ${\medmath{3.6778+1.8126i}}$ & ${\medmath{7.7206+1.5315i}}$\\
                       & {\footnotesize WKB:} & -- & -- & -- & -- & -- & -- \\ \hline
\multirow{2}{*}{$4$} & {\footnotesize Leaver:} & ${\medmath{1.3824+1.6186i}}$ & ${\medmath{1.9458+2.0143i}}$ & ${\medmath{2.5611+2.3806i}}$ & ${\medmath{3.3298+2.6388i}}$ & ${\medmath{3.9017+2.4707i}}$& ${\medmath{7.8794+2.0576i}}$\\
                       & {\footnotesize WKB:} & -- & -- & -- & -- & -- & -- \\ \hline
\multirow{2}{*}{$5$}& {\footnotesize Leaver:}  & ${\medmath{1.6012+1.8188i}}$ & ${\medmath{2.1982+2.1838i}}$ & ${\medmath{2.8820+2.5640i}}$ & ${\medmath{3.6142+2.9372i}}$ & ${\medmath{4.1703+3.0841i}}$ &${\medmath{8.0509+2.6145i}}$\\
                       & {\footnotesize WKB:} & -- & -- & -- & -- & -- & -- \\ \hline
\multirow{2}{*}{$6$} & {\footnotesize Leaver:} & ${\medmath{1.8610+2.0158i}}$ & ${\medmath{2.4649+2.3893i}}$ & ${\medmath{3.1606+2.7566i}}$ & ${\medmath{3.9260+3.1271i}}$ & ${\medmath{4.4112+3.2877i}}$ &${\medmath{8.2341+3.1993i}}$ \\
                       & {\footnotesize WKB:} & -- & -- & -- & -- & -- & -- \\ \hline
\multirow{2}{*}{$7$} & {\footnotesize Leaver:} & ${\medmath{2.1249+2.2063i}}$ & ${\medmath{2.7528+2.5864i}}$ & ${\medmath{3.4626+2.9606i}}$ & ${\medmath{4.2388+3.3311i}}$ & ${\medmath{4.7421+3.4980i}}$& ${\medmath{8.4290+3.8093i}}$\\
                       & {\footnotesize WKB:} & -- & -- & -- & -- & -- & -- \\ \hline
\multirow{2}{*}{$8$} & {\footnotesize Leaver:} & ${\medmath{2.4010+2.4033i}}$ & ${\medmath{3.0483+2.7844i}}$ & ${\medmath{3.7753+3.1607i}}$ & ${\medmath{4.5659+3.5341i}}$ & ${\medmath{5.0700+3.7007i}}$& ${\medmath{8.6360+4.4421i}}$\\
                       & {\footnotesize WKB:} & -- & -- & -- & -- & -- & --\\ \hline
\multirow{2}{*}{$9$}  & {\footnotesize Leaver:}  & ${\medmath{2.6906+2.6000i}}$ & ${\medmath{3.3558+2.9838i}}$ & ${\medmath{4.0981+3.3620i}}$ & ${\medmath{4.9025+3.7369i}}$ & ${\medmath{5.4096+3.9055i}}$ & ${\medmath{8.8784+5.0665i}}$ \\
                       & {\footnotesize WKB:} & -- & -- & -- & -- & -- & -- \\ \hline
\end{tabular}

\caption{The ten first QNM frequencies, $\omega_N$ for the topological star of the first kind ($r_\text{B} = 13/14$ and $r_\text{S}=15/28$), for $\ell$ from 0 to 4 and 10. The top and bottom lines in each cell are the values obtained from Leaver's method and WKB respectively. The agreement between both values is shown in Fig. \ref{fig:Percent_error_topstar1}. The black hole modes are starred, and the highly damped modes which cannot be accessed by our WKB methods are in plain text.
}
\label{tab:QNMs_high_damped}
\end{table*}

\begin{figure}[t]

\centering
\includegraphics[width=0.45\textwidth]{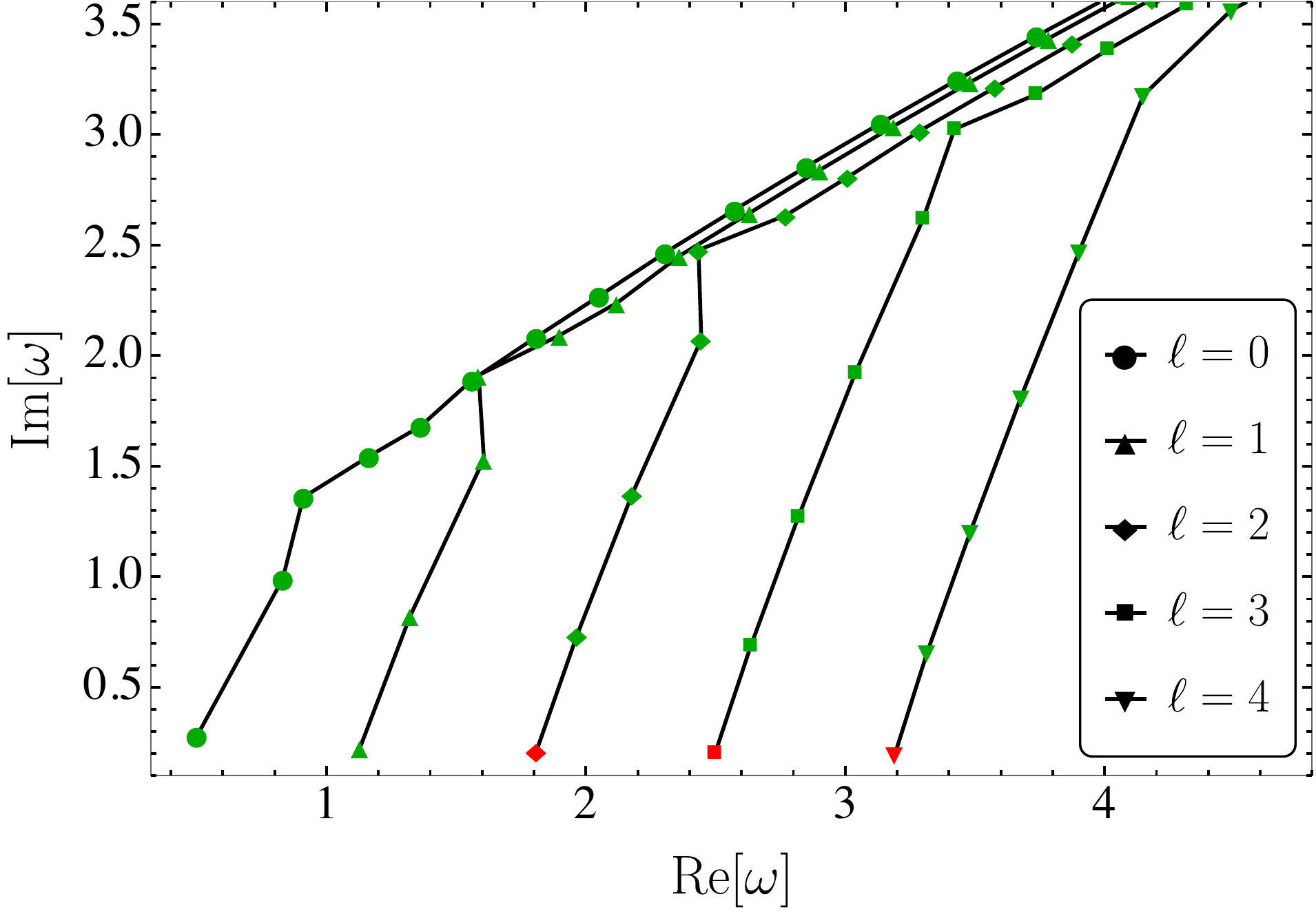}
\caption{The first few QNMs of the topological star of the first kind, with values listed in Table~\ref{tab:QNMs_high_damped}, and mass normalized to $M=1/2$. The black hole modes are shown in red, and the highly damped modes in green. }
\label{fig:QNMs_topstar1}

\end{figure}

\begin{figure}[t]

\centering
\includegraphics[width=0.45\textwidth]{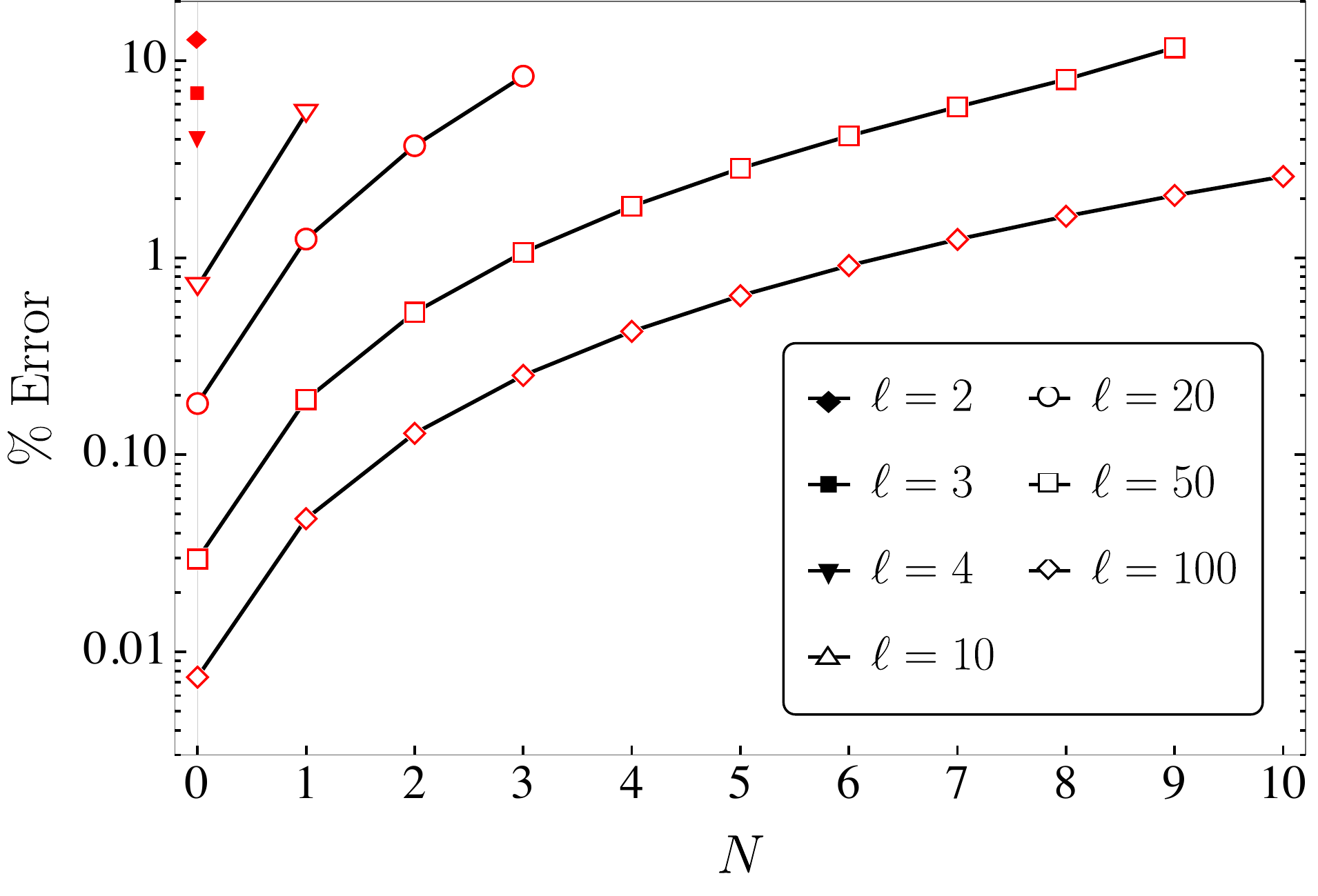}
\caption{Percent error between the WKB and Leaver frequencies of Table \ref{tab:QNMs_high_damped} and extended to higher $\ell$ values, for the topological star of the first kind.}
\label{fig:Percent_error_topstar1}

\end{figure}

We now consider the illustrative example of a topological star of the first kind with $r_\text{B}=13/14$ and $r_\text{S}=15/28$. The first few modes for $\ell=0,\dots,4$ are plotted in Fig.~\ref{fig:QNMs_topstar1}. In Table~\ref{tab:QNMs_high_damped} we list the QNM frequencies for $\ell=0,\dots,4$ as well as those for $\ell=10$ case, along with their WKB counterparts (when they exist). The relative difference between WKB and Leaver's frequencies is plotted in Fig.~\ref{fig:Percent_error_topstar1}. We classify the modes using the same conventions as in the previous section.

The modes with starred values in the table are the slowly damped modes of Eq.~\eqref{eq:WKBQNMTS1}, derived using our WKB method in Sec.~\ref{sec:WKBTS1}. Those are black hole-like modes localized at the unstable photon sphere and governed by the scattering properties there. Figure~\ref{fig:Percent_error_topstar1} shows that errors from the WKB approximation are below 5$\%$ for $\ell \geq 4$, and that the accuracy improves as $\ell$ increases. Note that the error is slightly larger for this WKB method than for the previous one. This is probably due to the two layers of approximation used: the WKB approximation and the approximation of the potential barrier by its asymptotic form. The latter approximation also causes the accuracy to decrease with $N$ as the barrier becomes larger.

Leaver's method allows us to compute several QNM frequencies in the highly damped QNM tower. They have similar properties as for the topological star of the second kind: both real and imaginary part increase with $N$, unlike the highly damped modes of a black hole with the same photon sphere.

\subsubsection{Direct comparison with a black hole}
\label{sec:BHcomp}

So far, we have indirectly compared topological stars to black holes by identifying modes that are associated with the outer unstable photon sphere.
We can push this comparison further by deriving explicitly the spectrum of the charged static black objects of the five-dimensional theory. They are charged black strings that lead to magnetic black holes from a four-dimensional perspective, given by the same metric and field as Eq.~\eqref{eq:met&GF} but with $r_\text{S} \geq r_\text{B}$~\cite{Bah:2020ogh}. However, these solutions do not exist in the same mass and charge ranges as stable topological stars. Therefore, a topological star cannot be easily compared with a black hole of the same mass and charge. This makes the comparison of their QNMs more problematic (the QNMs of the magnetic black string have been studied independently in Ref.~\cite{Guo:2022rms}).

Nevertheless, near-extremal topological stars of the second kind, i.e. $r_\text{B} \gtrsim r_\text{S}$, can be heuristically compared to a near-extremal black string of the same mass and almost the same charge with $r_\text{S} \gtrsim r_\text{B}$. In Table \ref{tab:blackstring_modes}, we compute the first QNMs of the near extremal black string, with $r_\text{S}=0.67$ and $r_\text{B}=0.66$ ($M=1/2$ and $Q=1.15$) and compare them with the QNMs of the near-extremal topological star studied in Sec.~\ref{sec:TS2Leaver}. Details of the derivation using Leaver's method can be found in Appendix~\ref{app:black_string_leaver}.

Table~\ref{tab:blackstring_modes} confirms the findings of the WKB method. The real parts of the black string QNMs are very similar to those we have identified as the ``black hole'' modes of the topological star. Remarkably, the agreement is quite good even for small $\ell$, where the WKB approximation should not be very accurate, and it improves as $\ell$ increases. Moreover, the QNM imaginary parts are much larger for the black string, as expected. The time-domain signal should be a very similar signal in the short term, but have a longer damping for the topological star, due to the cavity effect produced by the pair of photon spheres.

The table also shows that the black string and the topological star have  different highly damped modes (values in plain text): the real part of the QNM frequencies of the black string decreases with $N$, unlike the topological star. 

\begin{table*}[ht]
\setlength{\tabcolsep}{0.4em}
\begin{tabular}{| l l | c | c | c | c | c |}
\hline \hline  
\makecell{$N$}& & \makecell{$\ell=0$}& \makecell{$\ell=1$} & \makecell{$\ell=2$} & \makecell{$\ell=3$} & \makecell{$\ell=4$} \\ \hline \hline
\multirow{1}{*}{$0$} & {\footnotesize Leaver:} & ${\medmath{0.3590+0.2162i}}$ & ${*~\medmath{0.9087+0.1801i}~*}$ & ${*~\medmath{1.4826+0.1749i}~*}$ & ${*~\medmath{2.0624+0.1812i}~*}$ & ${*~\medmath{2.7956+0.5989i}~*}$ \\ 
& {\footnotesize BH modes of TS:} & -- & ${*~\medmath{0.9112+0.0150i}~*}$ & ${*~\medmath{1.3965+0.0036i}~*}$ & ${*~\medmath{2.0611 +0.0120i}~*}$ & ${*~\medmath{2.7202+0.0275i}~*}$ \\ \hline

\multirow{1}{*}{$1$} & {\footnotesize Leaver:} & ${\medmath{0.3279+0.8789i}}$ & ${\medmath{0.8616+0.6683i}}$ & ${\medmath{1.4616+0.6149i}}$ & ${\medmath{2.0760+0.5881i}}$ & ${*~\medmath{2.5849+0.8622i}~*}$ \\ 
& {\footnotesize BH modes of TS:} & -- & -- &-- & -- & ${*~\medmath{2.5520+0.0030i}~*}$\\ \hline

\multirow{1}{*}{$2$} & {\footnotesize Leaver:} & ${\medmath{0.2131+1.6122i}}$ & ${\medmath{0.6528+1.2539i}}$ & ${\medmath{1.3037+1.1163i}}$ & ${\medmath{1.9434+1.0323i}}$ & ${\medmath{2.3282+1.3698i}}$ \\
& {\footnotesize BH modes of TS:} & -- & -- & -- & -- & --\\ \hline

\multirow{1}{*}{$3$} & {\footnotesize Leaver:} & ${\medmath{0.0550+2.2270i}}$ & ${\medmath{0.3751+1.8876i}}$ & ${\medmath{1.0241+1.6465i}}$ & ${\medmath{1.6992+1.5220i}}$ & ${\medmath{2.0340+1.8720i}}$ \\ 
& {\footnotesize BH modes of TS:} & -- & -- & -- & -- & -- \\ \hline

\end{tabular}

\caption{The first modes of the near-extremal black string with $r_\text{S}=0.67$ and $r_\text{B}=0.66$ ($Q=1.1518$), compared to the ``black hole'' modes of the topological stars given in Table~\ref{tab:QNMs_and_WKB} (starred).}
\label{tab:blackstring_modes}
\end{table*}

\section{Conclusions}\label{sec:conclusion}

In this paper, we have shown that the spectrum of smooth topological solitons contains a sub-class of ``black hole-like'' modes, because both solutions have an outer photon shell with the same scattering properties. 
To prove this claim, we have computed the QNMs of certain classes of topological stars that serve as interesting prototypes of topological solitons. The linear perturbation equations for these topological stars are separable, and their QNM spectrum can be computed analytically. 

First, we have shown that topological stars that have a single unstable photon sphere and no horizon have black hole-like QNMs. This result indicates that the oscillation frequencies of black holes can be well approximated even in the absence of a horizon. 
Second, we considered topological stars with two photon spheres (an inner, stable orbit and an outer, unstable orbit). We have highlighted two important features in their spectrum. First, the fundamental modes are localized at the inner orbit and they are determined by the scattering properties there. Second, black hole-like modes are present in the spectrum, but the imaginary parts of their frequencies differ from those of a black hole due to a ``cavity effect'' produced by the photon spheres. 
This leads to QNMs having longer damping than what one would expect for black holes. To our knowledge, the damping difference and the cavity effect have not been discussed in the literature, and they could lead to interesting phenomenology. 

We believe that these results also apply to any smooth horizonless geometry and that they depend only on the generic stability properties of photon spheres. It would be interesting to develop this generalization in future work. Ultimately, we would like to apply a similar argument to more astrophysically relevant topological solitons, such as Schwarzschild topological solitons or bubble bag ends~\cite{Bah:2022yji}. Although these geometries are axially symmetric, which complicates the analysis considerably, the structure of their photon spheres is relatively similar to that of a topological star of the second kind, with a shadow-like photon sphere surrounding intertwined stable orbits. Therefore, these geometries should have black hole-like modes with damping differences  determined by the outer photon sphere, as well as a more complex spectrum of microstructure modes that could induce chaotic ``echoes'' at late time. We would like to address this problem (and to better understand the linear stability of these solutions) in the near future. 

In this paper, we have focused on scalar perturbations to simplify the discussion. Ultimately, we would like to extend our work to gravitational perturbations. 

Finally, we would like to address the physics of nonlinear modes. Recently, nonlinear spectroscopy of black holes or black-hole mimickers has attracted substantial attention. 
First, it has been shown that nonlinear effects have a significant impact on the ringdown signal of a black hole merger~\cite{Cheung:2022rbm,*Baibhav:2023clw}. 
Second, in four-dimensional general relativity, stable photon orbits are suspected to trigger nonlinear instabilities, and this has been confirmed recently using numerical evolutions of boson stars~\cite{Cunha:2022gde,*Keir:2014oka}. These results  call into question the relevance of four-dimensional exotic ultra-compact objects to describe black hole-like geometries. 
However, they only apply to general relativity in four dimensions. Indeed, these nonlinear instabilities are intimately related to the no-soliton theorem in four dimensions~\cite{serini1918euclideita,*breitenlohner19884}, which implies that there can be two possible final states for ultracompact exotic objects: migration to non-ultracompact configurations or total gravitational collapse. 

Theories of gravity with extra-compact dimensions, that emerge naturally from string theory, evade this no-soliton theorem, so the fate of nonlinear instabilities may be different.  It is expected that wide classes of topological solitons may exist and comprise a dense set of states in quantum gravity.  These may admit nontrivial quantum tunneling not just to one another, but to the even larger set of generic quantum states with the same quantum numbers.  From this perspective, the nonlinear instability is ``more of a feature than a bug'' for spacetimes with extra compact dimensions in quantum gravity. The final state might be different from a black hole.  These expectations, while motivated by the vast constructions of topological solitons, are subject to debate and deserve further study.  A fundamental avenue of research in quantum gravity is to precisely understand the mechanism for such a tunneling process.  We can highlight various perspectives on why these are interesting and nontrivial expectations.

First, we can consider the illustrative case of supersymmetric solitons. The largest family of Einstein equations to date has been derived in Ref.~\cite{Heidmann:2019xrd} and corresponds to smooth horizonless geometries, not given in terms of free parameters but of arbitrary functions of three spacetime coordinates. These functions give an arbitrary shape to smooth wiggles of spacetime added on a bubbling geometry. Moreover, these wiggles correspond to momentum modes that have backreacted on the geometry, not by collapsing the solution into a black hole, but into a smooth wiggly topological microstructure. It would be interesting to see if such a mechanism can be reproduced with non-supersymmetric topological solitons, especially with topological stars~\cite{Bah:2020ogh,Bah:2020pdz,Bah:2022pdn}. This would show that these topological states of gravity can absorb energy by forming spacetime wiggles without collapsing into a black hole. Such a mechanism cannot exist in four-dimensional general relativity, and requires new geometric degrees of freedom of classical theories of gravity that emerge naturally from string theory. 

Second, the derivation in Ref.~\cite{Bah:2021irr} has shown that topological stars are meta-stable states of the five-dimensional Einstein-Maxwell theory. They are therefore viable thermodynamic phases of the theory that should be put on an equal footing with the black hole phase.  Moreover, Hawking-Page phase transitions exist between the topological star phase and the black hole phase. This suggests that such topological states could be nucleated in a physical process, and objects other than black holes may be generated.  Such a nucleation will require quantum effects triggered, potentially, by the nonlinear instabilities. 

As a final argument, we can use the nature of topological solitons as states of quantum gravity. These states correspond to coherent bound states of strings and branes that have produced a geometric transition to smooth horizonless geometries. From a quantum perspective, perturbations of coherent states should lead to their decay into less coherent states in the phase space. If the latter have the same constituents of branes and strings, they would not necessarily admit geometric descriptions as gravitational solitons, but rather be described by inherently fuzzy and quantum states of quantum gravity.

We plan to explore these topics in the future and understand which of these scenarios are physically plausible.  While exciting, this research program will require a conceptual new understanding of the dynamics of topological solitons, and their more generic quantum cousins.  This is key to determining the final states of the nonlinear instability.

%%%%%%%%%%%%%%%%%%%%%%%%%%%%%%%%%%%%%%%%%%%%%%%%%%%%%%%%%%%%%%%%%%%%%%%%%%%%%%%%%%%%%%%
\acknowledgments 
While this work was in preparation, we became aware of related work by Massimo Bianchi, Giorgio Di Russo, Alfredo Grillo, Francisco Morales and Giuseppe Sudano~\cite{RomeWork}. 

We thank Sebastian Schulz, Hannah Tillim and Peter Weck for interesting discussions.
The work of P.H. and IB is supported by NSF grant PHY-2112699.   The work of IB is also supported in part by the Simons Collaboration on Global Categorical Symmetries.
E.B. and N.S. are supported by NSF Grants No. AST-2006538, PHY-2207502, PHY-090003 and PHY-20043, and NASA Grants No. 19-ATP19-0051, 20-LPS20-0011 and 21-ATP21-0010.

%%%%%%%%%%%%%%%%%%%%%%%%%%%%%%%%%%%%%%%%%%%%%%%%%%%%%%%%%%%%%%%%%%%%%%%%%%%%%%%%%%%%%%%

\appendix

\section{WKB methods}
\label{app:WKBGen}

We have argued that the potential for scalar perturbations in topological stars has up to two roots for $r> r_\text{B}$ when all parameters are real-valued. The WKB approximation is used to estimate the spectrum of slowly damped QNMs that have $\text{Im}(\omega)\ll \text{Re}(\omega)$. We detail the derivation when the potential has one or two roots.

\subsection{The two-root scenario}
\label{app:WKB}

We review the derivation of Ref.~\cite{Bena:2020yii} and adapt it to our type of potential. 

We consider a potential for which the imaginary part can be neglected at first and that has two zeroes. We assume that it depends on a generic variable $r^*\in \mathbb{R}$,  such that
$V(r^*) \to -\omega^2$ at large $r^*$ with $\text{Re}(\omega)>0$.  However,  unlike Ref.~\cite{Bena:2020yii},  here the potential goes to zero at the other boundary $r^*\to -\infty$ as
\begin{equation}
V(r^*) \underset{r^*\to -\infty}{\sim} -a \,e^{b\,r^*}\,,\qquad a,b>0.
\label{eq:AppPotAsym}
\end{equation}
A typical potential is shown in Fig.~\ref{fig:Pot}. The scalar wave is approximated by the WKB waveform
\begin{equation}
K(r^*)=
\left\{
\arraycolsep=0.1pt\def\arraystretch{1.8}
\begin{array}{rll}
& \dfrac{1}{|V(r^*)|^{\frac{1}{4}}} \left[D^\text{I}_+ \,e^{i \mathcal{I}(-\infty,r^*)}+ D^\text{I}_-  \,e^{-i \mathcal{I}(-\infty,r^*)} \right] ,\\
& \dfrac{1}{|V(r^*)|^{\frac{1}{4}}} \left[D^\text{II}_+ \,e^{\mathcal{I}(r_0^*,r^*)}+ D^\text{II}_-  \,e^{- \mathcal{I}(r_0^*,r^*)} \right],\\
& \dfrac{1}{|V(r^*)|^{\frac{1}{4}}} \left[D^\text{III}_+ \,e^{i \mathcal{I}(r_1^*,r^*)}+ D^\text{III}_-  \,e^{-i \mathcal{I}(r_1^*,r^*)} \right] , 
\end{array}
\right. 
\label{eq:wavefunctionWKB}
\end{equation}
where the first,  second,  and third lines correspond to the three zones ($r^*< r_0^*$,  $r_0^*< r^*<r_1^*$, and $r_1^*< r^*$, respectively),  and we have defined
\begin{equation}
\mathcal{I}(a,b)\, \equiv\,\int_{a}^b |V(r^*)|^{\frac{1}{2}}dr^*.
\end{equation}
The constants $D^\text{I}_\pm$,  $D^\text{II}_\pm$ and $D^\text{III}_\pm$ are related to each other by the junction rules using Airy function matching at the turning points:
\begin{align}
\label{eq:Appconnection1}
\begin{pmatrix} 
D^\text{II}_+ \\ 
D^\text{II}_-
\end{pmatrix} &~=~\begin{pmatrix} 
e^{i(\Theta+\frac{\pi}{4})} & e^{-i(\Theta+\frac{\pi}{4})} \\ 
\tfrac{1}{2}\,e^{i(\Theta-\frac{\pi}{4})} & \tfrac{1}{2}\,e^{-i(\Theta-\frac{\pi}{4})}
\end{pmatrix}  \begin{pmatrix} 
D^\text{I}_+ \\ 
D^\text{I}_-
\end{pmatrix}, \nonumber\\
\begin{pmatrix} 
D^\text{III}_+ \\ 
D^\text{III}_-
\end{pmatrix} &~=~\begin{pmatrix} 
e^{T-i\frac{\pi}{4}} & \tfrac{1}{2}\,e^{-T+i\frac{\pi}{4}} \\ 
e^{T+i\frac{\pi}{4}} & \tfrac{1}{2}\,e^{-T-i\frac{\pi}{4}}
\end{pmatrix}  \begin{pmatrix} 
D^\text{II}_+ \\ 
D^\text{II}_-
\end{pmatrix},
\end{align}
where $\Theta$ and $T$ have been defined in \eqref{eq:Theta&Tdef1}.  

At large $r^*$,  the WKB waveform gives
\begin{equation}
K(r^*) \propto D^\text{III}_+  \,e^{i \,\omega r^*} + D^\text{III}_-  \,e^{-i \,\omega r^*}\,.
\end{equation}
Imposing the outgoing boundary condition at $r^*=\infty$ as in Eq.~\eqref{eq:OutgoingBC} therefore requires 
\begin{equation}
D^\text{III}_+\=0.
\label{eq:AppCondAsymp}
\end{equation} 

The boundary condition at the origin $r^*=-\infty$ is more subtle since it is where the potential vanishes.  We cannot directly use the WKB waveform to constrain the constants $ D^\text{I}_\pm$.  We follow the usual WKB procedure at a zero, which is to solve for the potential locally at $r^*\sim -\infty$ using Eq.~\eqref{eq:AppPotAsym}; impose the boundary condition on the local wave obtained in this way; and then take the asymptotic expansion of this wave at $r^\star\to \infty$, and match it with the asymptotic expansion of the WKB waveform at $r^\star\to -\infty$.

First, the solutions of the Schr\"odinger equation with the potential \eqref{eq:AppPotAsym} are given by
\begin{equation}
K_{-\infty}(r^*) \= c_1 \,  J_0\left( \frac{2\sqrt{a}}{b}\,e^{\frac{br^*}{2}}\right) + c_2\,  Y_0\left( \frac{2\sqrt{a}}{b}\,e^{\frac{br^*}{2}}\right),
\end{equation}
where $J_0$ and $Y_0$ are the Bessel functions of order $0$ of the first and second kind.  By studying the behavior of both functions asymptotically,  one can check that only $J_0$ satisfies the boundary condition \eqref{eq:RegOrigin}, such that $J_0(r^*) \sim 1 + \mathcal{O}(e^{b r^*})$.  Thus we consider $c_2=0$.  

Moreover,  at large $r^*$ we have
\begin{align}
J_0\left( \frac{2\sqrt{a}e^{\frac{br^*}{2}}}{b}\right)  \sim \frac{\sqrt{b}\,e^{-\frac{b r^*}{2}}}{2\sqrt{\pi\,\sqrt{a}}}& \Biggl(\exp\left[i \left( \frac{\pi}{4}-\frac{2\sqrt{a}e^{\frac{br^*}{2}}}{b}\right) \right] \nonumber \\
& \hspace{-0.5cm} + \exp\left[i \left(\frac{2\sqrt{a}e^{\frac{br^*}{2}}}{b}- \frac{\pi}{4}\right) \right]  \Biggr).
\end{align}

Now, expand the second line of Eq.~\eqref{eq:wavefunctionWKB} at $r^*\to -\infty$:
\begin{equation}
K(r^*) \sim \frac{e^{-\frac{b r^*}{2}}}{a^\frac{1}{4}} \left(D^\text{I}_\pm \, \exp\left[\pm \frac{2i\sqrt{a}e^{\frac{br^*}{2}}}{b} \right] \right).
\end{equation}
By matching to the Bessel function,  we find that the boundary condition at the origin requires the WKB integration constant to satisfy:
\begin{equation}
\begin{split}
D_-^\text{I} &\=i\,D_+^\text{I} \,.
\end{split}
\label{eq:AppCondOrigin}
\end{equation}

Finally,  we can use the junction rules \eqref{eq:Appconnection1} to determine when the two boundary conditions \eqref{eq:AppCondAsymp} and \eqref{eq:AppCondOrigin} are compatible.  We find the condition:
\begin{equation}
\begin{split}
\cos \Theta+ i\, \frac{e^{-2T}}{4}\,\sin \Theta &\,=\,0 \,.
\end{split}
\label{eq:AppWKBSpec}
\end{equation}

\subsection{The single-root scenario}
\label{app:WKB2}

Consider now a potential with a single root such that
\begin{equation}
V(r^*) \underset{r^*\to \infty}{\sim} -\omega^2\,,\quad V(r^*) \underset{r^*\to -\infty}{\sim} a \,e^{b\,r^*}(1-c \,e^{b\,r^*})\,,
\label{eq:AppPotAsym1zero}
\end{equation}
where $a,b,c$ are positive real constants. A typical potential is shown in Fig.~\ref{fig:Pot2}. 

Since the potential consists of a single potential barrier, one can think that the WKB approximation cannot capture QNMs. Indeed, the same derivation as the previous section will not lead to any spectrum.  The barrier induces a scale factor of $e^{-T}$ between the transmitted and reflected amplitudes, while having purely outgoing modes asymptotically requires both amplitudes to be of the same order.   

However, this also occurs for black hole QNMs~\cite{Schutz:1985km,Iyer:1986np}. QNMs have been derived when the barrier is small such that $e^{-T}$ is of order one. For black holes, this is done by connecting the WKB waveform at the horizon with the waveform at the asymptotics through a small barrier approximated by a quadratic potential. 

The situation is different for our smooth geometries, but similar in spirit. We do not have a horizon region, and the potential goes to zero at the interior boundary $r^* \to -\infty$. However, by assuming that the barrier is small, the turning point, $r^*_0$, is sufficiently close to the boundary such that the whole barrier is well-approximated by the asymptotic form of the potential \eqref{eq:AppPotAsym1zero}. This requires
\begin{equation}
    e^{b r_0^*} \,\sim\, \frac{1}{c} \,\ll\, 1\quad \Longleftrightarrow \quad  V(-b^{-1}\log c) \approx 0.
\end{equation}

We therefore locally solve the equation at the barrier,
\begin{equation}
    \partial_{r^*}^2 K - a \,e^{b\,r^*}(1-c \,e^{b\,r^*}) \, K \= 0\,, 
    \label{eq:AppBondEq}
\end{equation}
take the branch that does not diverge as $r^*\to -\infty$, and match it to the outgoing wave outside the barrier.

Solutions of Eq.~\eqref{eq:AppBondEq} are given by
\begin{align}
    K \= & e^{i \mu \,e^{br^*}}\, \Bigl( A \,\, {}_1 F_1 (\nu+1 ,1, -2 \mu \,i\, e^{b r^*}) \nonumber \\
    &\hspace{1.8cm} + B\,\,U(\nu+1 ,1, -2 \mu \,i\, e^{b r^*})\Bigr) \,, \\
    \nu \equi & -\frac{1}{2} + \frac{i\,\sqrt{a}}{2 b \sqrt{c}}\,,\qquad \mu \equi \frac{\sqrt{a c}}{b}\,,\nonumber 
\end{align}
where $(A,B)$ are integration constants, and ${}_1 F_1 $ and $U$ are the confluent hypergeometric functions. The asymptotics of $K$ are given by
\begin{align}
    K &\underset{r^*\to -\infty}{\sim} B \frac{b}{\Gamma(\nu+1)} \,r^* + \mathcal{O}(1)\,, \nonumber\\
    K &\underset{r^*\to \infty}{\sim} \frac{e^{i \mu e^{b r^*}}}{(2 i \mu e^{b r^*})^{\nu+1}} \left( \frac{A}{\Gamma(-\nu)} - (-1)^{-\nu} B\right) \\\
    &\hspace{1cm} +\frac{e^{-i \mu e^{b r^*}}}{(2 i \mu e^{b r^*})^{-\nu}} \frac{A}{\Gamma(\nu+1)} \,.\nonumber
\end{align}
First, smoothness at $r^*\to \infty$ requires $B=0$, while the outgoing part of the wave is the one proportional to $e^{-i \mu e^{br^*}}$ at a large distance. The QNM condition therefore requires
\begin{equation}
   \Gamma(-\nu) \= \infty \quad \Leftrightarrow \quad i  \frac{\sqrt{a}}{2 b \sqrt{c}} \= N + \frac{1}{2} \,, 
\end{equation}
where $N$ is a positive integer. Since the maximum of the potential is well approximated by its asymptotic form, one can relate $a,b,c$ to the properties of the potential there. We find
\begin{equation}
 i\,\frac{\sqrt{2}\, V(r^*_\text{max})}{\sqrt{|V''(r^*_\text{max})|}} \= N + \frac{1}{2} \,, 
\end{equation}
The formula has a strong similarity with the WKB formula for black holes~\cite{Schutz:1985km,Iyer:1986np}, while the boundary condition applied at the origin of spacetime is very different.

\section{Derivation of the barrier integral}
\label{App:ImaginaryCorrection}

The topological stars of the second kind have almost-trapped scalar modes for which the imaginary part of the frequencies are given by \eqref{eq:WKBspectrum}.  One needs to derive the barrier integral $T$ \eqref{eq:Theta&Tdef1} for that purpose.  The integral can only be expressed in terms of elliptic functions. 
\begin{widetext}
We obtain
\begin{align}
&T \=  \frac{\omega}{\sqrt{(r_0+2 r_1)(r_0-r_\text{B})}\,(r_\text{B}-r_\text{S})} \Biggl[ -(r_0+2 r_1)(r_0-r_\text{B})(r_\text{B}-r_\text{S})\,E\left(\frac{(r_0-r_1)(r_0+r_1+r_\text{B})}{(r_0+2 r_1)(r_0-r_\text{B})} \right) \nonumber\\
&\hspace{1cm} - (r_1-r_\text{B})(r_0-r_\text{B})(2 r_0+2 r_1+r_\text{B}+r_\text{S}) \, K\left(\frac{(r_0-r_1)(r_0+r_1+r_\text{B})}{(r_0+2 r_1)(r_0-r_\text{B})} \right) \nonumber \\
&\hspace{1cm} - (r_1-r_\text{B})(r_\text{B}-r_\text{S})(r_\text{B}+2r_\text{S}) \, \Pi\left(\frac{r_0-r_1}{r_0-r_\text{B}}\,,\,\frac{(r_0-r_1)(r_0+r_1+r_\text{B})}{(r_0+2 r_1)(r_0-r_\text{B})} \right)\\
&\hspace{1cm} + 2(r_1-r_\text{B})(r_0-r_\text{S})(r_0+r_1+r_\text{S})\,\, \Pi\left(\frac{(r_0-r_1)(r_\text{B}-r_\text{S})}{(r_0-r_\text{B})(r_1-r_\text{S})}\,,\,\frac{(r_0-r_1)(r_0+r_1+r_\text{B})}{(r_0+2 r_1)(r_0-r_\text{B})} \right)\Biggr] \nonumber 
\end{align}
\end{widetext}
where $E$,  $K$ and $\Pi$ are the complete elliptic integral,  the complete elliptic integral of the first kind, and the complete elliptic integral of the third kind respectively.  Moreover,  $r_0$ and $r_1$ are the positive roots of the potential:
\begin{equation}
\begin{split}
   & \medmath{r_k = \frac{2\sqrt{\ell(\ell+1)}}{\omega \sqrt{3}} \cos \Biggl[ \frac{1}{3} \arccos \left(-\frac{3\sqrt{3}\omega(r_\text{S}(\ell(\ell+1)+1)-r_\text{B})}{2(\ell(\ell+1))^{3/2}} \right)}\\
    & \hspace{3cm}\medmath{+\frac{2\pi}{3}(k-1)  \Biggr].}
\end{split}
\nonumber
\end{equation}

\section{Leaver's method}
\label{app:Leaver}

In this section, we give more details on the application of Leaver's method to topological stars and black strings. 

\subsection{Topological stars}
\label{appendix:numerical_methodology}

By inserting the series expansion \eqref{eq:Leaver_series} in the wave equation \eqref{eq:PotGen}, we obtain the recurrence relation \eqref{eq:leaver_coeffs} for the coefficients $a_n$, where $\alpha_n$, $\beta_n$, $\gamma_n$, and $\delta_n$ are explicitly given by
\begin{equation}
\label{eq:leaver_coeffs2}
\begin{aligned}
\alpha_n =& (n+1)^2 \\
\beta_n =& -(1+\ell +\ell^2 +2n+3n^2)\\ &- (1+2n)(r_\text{B}-r_\text{S}) i \omega + \frac{r_\text{B}^3 \omega^2}{r_\text{B}-r_\text{S}} \\
\gamma_n =&1+\ell +\ell^2 -2n+3n^2 \\& + (2n-1)(r_\text{B}-r_\text{S}) i \omega + \frac{r_\text{S}^2(r_\text{S}-3r_\text{B}) \omega^2}{r_\text{B}-r_\text{S}} \\
\delta_n =& -(n-1)^2+ \frac{r_\text{S}^3 \omega^2}{r_\text{B}-r_\text{S}}.
\end{aligned}
\end{equation}
We reduce this four-term relation to a three-term recurrence relation of the form~\eqref{eq:leaver_coeffs1} using the Gaussian elimination 
\begin{equation}
\label{eq:primed_leaver_coeffs}
\alpha'_n = \alpha_n, ~~~ \beta'_n = \beta_n - \frac{\alpha'_{n-1}\delta_n}{\gamma'_{n-1}}, ~~~ \gamma'_n = \gamma_n - \frac{\beta'_{n-1}\delta_n}{\gamma'_{n-1}}.
\end{equation}
The condition for having a convergent series for the $a_n$ gives the QNM condition \eqref{eq:ContFrac} and its inverted forms \eqref{eq:Leaver_inversions}, which we solve numerically to find the frequency $\omega$. 

The inverted forms are useful for two reasons. First, numerical searches for the $N^{\text{th}}$ QNM are more stable when we look for roots of the $N^{\text{th}}$ inversion~\cite{Leaver:1985, Leaver:1990zz}. Second, the two equations \eqref{eq:ContFrac} and \eqref{eq:Leaver_inversions} have the same solutions in principle, but in practice one may find spurious modes due to the truncation of the infinite continued fractions when solved numerically. To confirm that the roots indeed correspond to QNM frequencies, we vary the number of inversions in Eq.~\eqref{eq:Leaver_inversions} and check that the roots remain stable under these inversions. 

We also calculate the ``remainder'' of the truncated continued fraction in Eq. \eqref{eq:ContFrac}, as first introduced by Nollert~\cite{Nollert:1993zz}. To solve numerically the continued fractions, we truncate them at some large but finite value, say $n$. Numerical convergence can be improved if we make a specific choice for the ``rest'' of the continued fraction. The remainder $R_n$ which corresponds to the truncated part of the fraction is defined by
\begin{equation}
\label{eq:Nollert_remainder}
R_n=\frac{\gamma'_{n+1}}{\beta'_{n+1}-\alpha'_{n+1}R_{n+1}}.
\end{equation}
Expanding $R_n$ in a series of the form
\begin{equation}
\label{eq:Nollert_series}
R_n=\sum_{k=0}^{\infty}C_k n^{-k/2},
\end{equation} 
we can find an approximation of the remainder as $n\rightarrow \infty$. The first few coefficients are $C_0=-1$, $C_1=\pm \sqrt{2 i \omega (r_\text{B}-r_\text{S})}$, $C_2=\frac{1}{4}(3-(5r_\text{B}+2r_\text{S})i\omega)$, and $C_3=[3/32+\ell(\ell+1)/2-\omega(37r_\text{B}^2\omega+r_\text{B}(76i-20r_\text{S}\omega)+4r_\text{S}(-22i+r_\text{S} \omega))]/C_1$. Following~\cite{Nollert:1993zz}, we pick the sign for $C_1$ such that $\text{Re}[C_1]>0$.

\subsection{Black strings}
\label{app:black_string_leaver}

In Sec.~\ref{sec:BHcomp}, we derived the first few modes of a near-extremal black string and compare them with those of a near-extremal topological star. The black string solution has the same metric as in Eq.~\eqref{eq:met&GF}, but with $r_\text{S} \geq r_\text{B}$, so that the $r=r_\text{S}$ locus corresponds to a horizon. This implies a different mode expansion and boundary conditions than those used for topological stars in Sec.~\ref{sec:ScalWave}.

First, we use the expansion 
\begin{equation}
\Phi_{\ell,m,\omega,p}(t,r,\theta,\phi,y) \,=\,  \frac{K(r)}{r-r_\text{B}} \, Y_{\ell}^{\,m}(\theta,\phi)\,e^{i\left(\omega t + p \frac{y}{R_y}\right)}\,,
\label{eq:FourierBlackString}
\end{equation}
where we have replaced the $(r-r_\text{S})^{-1}$ by $(r-r_\text{B})^{-1}$ to avoid divergences. The potential for massless scalars with no momentum along $y$ is now given by
\begin{align}
&\frac{r-r_\text{S}}{r-r_\text{B}}\, \partial_r \left( \frac{r-r_\text{S}}{r-r_\text{B}}\,\partial_r \,K\,\right)  - V(r) \,K \,=\, 0\,, \nonumber\\
&\medmath{V(r)\equiv \frac{\left(r-r_\text{S}\right)}{\left(r-r_\text{B} \right)^{4}} \left(r_\text{S}-r_\text{B} +\ell(\ell+1) \,(r-r_\text{B})-\omega^2\,r^3 \frac{r-r_\text{B}}{r-r_\text{S}}\right),}
\label{eq:Black_String_potential}
\end{align}
where the radial coordinate ranges from the horizon to spatial infinity, $r_\text{S}\leq r \leq \infty$. The equation can be recast into a Schrödinger form $\partial_{r^*}^2K-VK=0$ with a change of variable to a new tortoise-like coordinate:
\begin{equation}
\label{eq:tortoise_coord_black_string}
r^* \equiv r-r_\text{S} + (r_\text{S}-r_\text{B}) \log (r-r_\text{S}).
\end{equation}
In contrast to the topological star, the black string boundary conditions are the same as for a black hole: ingoing waves at the horizon, and outgoing waves at spatial infinity. Thus, we impose the following series expansion of $K(r)$ to apply Leaver's method:
\begin{equation}
\nn
\medmath{K(r)=e^{-i\omega r} (r-r_\text{B})^{-i \omega (r_\text{S}-r_\text{B})}\left( \frac{r-r_\text{S}}{r-r_\text{B}}\right)^{i \omega \sigma} \sum_{n=0}^\infty a_n \left( \frac{r-r_\text{S}}{r-r_\text{B}}\right)^n, }
\end{equation}
where $\sigma=\frac{r_\text{S}^{3/2}}{\sqrt{r_\text{S}-r_\text{B}}}$. Inserting the expansion into the radial equation \eqref{eq:Black_String_potential} yields a four-term recurrence relation for the $a_n$'s, similar to Eq. \eqref{eq:leaver_coeffs}. The coefficients in the relation are explicitly given by
\begin{equation}
\label{eq:leaver_coeffs_black_string}
\begin{aligned}
\alpha_n =& (n+1) \sigma ^4 (n+2 i \sigma \omega +1), \\
\beta_n =& ~\sigma^3 \omega^2(4r_\text{S}^3+3r_\text{S}^2\sigma+\sigma^3) \\ 
        & ~ -2i\sigma^2\omega((1+2n)r_\text{S}^3+(1+3n)\sigma^3) \\ & ~~ -\sigma^4(1+\ell +\ell^2+n(3n+2)),\\
\gamma_n =& ~\omega^2(3r_\text{S}^4\sigma^2-4r_\text{S}^6-6r_\text{S}^3\sigma^3-3r_\text{S}^2\sigma^4-2\sigma^6) \\
            &  ~+2i(3n-1)\sigma^2(r_\text{S}^3+\sigma^3)\omega \\ & ~ ~+(1+\ell+\ell^2+n(3n-2))\sigma^4,\\
\delta_n =& \left( (r_\text{S}^3+\sigma^3) \omega-i(n-1)\sigma^2 \right)^2.
\end{aligned}
\end{equation}
We then proceed as in Appendix \ref{appendix:numerical_methodology}, reducing the relation to three terms, and solving for the QNM frequencies numerically using Eqs.~\eqref{eq:ContFrac} and~\eqref{eq:Leaver_inversions}. To improve the convergence of the series expansion, we also compute the remainder coefficients, as in Eq.~\eqref{eq:Nollert_series}. The first few coefficients for the black string case are $C_0=-1$, $C_1=\pm \sqrt{2 i \omega(r_\text{S}-r_\text{B})}$, and $C_2=\frac{1}{4}(3+(r_\text{B}-r_\text{S})i\omega)$. Again, we pick the sign for $C_1$ so that $\text{Re}[C_1]>0$.

\bibliography{microstates}

\end{document}